\documentclass[%
aip,
amsmath,amssymb,
reprint,
]{revtex4-1}

\usepackage{graphicx}
\usepackage{dcolumn}
\usepackage{bm}
\usepackage{hyperref}
\usepackage[utf8]{inputenc}
\usepackage[T1]{fontenc}
\usepackage{mathptmx}
\usepackage{etoolbox}
\usepackage{braket}
\usepackage{float}
\usepackage{xcolor}
\providecommand{\revcolor}{black}
\newcommand{\rev}[1]{{\color{\revcolor}#1}}
\newenvironment{revblock}{\color{\revcolor}}{}

\makeatletter
\def\@email#1#2{%
\endgroup
\patchcmd{\titleblock@produce}
{\frontmatter@RRAPformat}
{\frontmatter@RRAPformat{\produce@RRAP{*#1\href{mailto:#2}{#2}}}\frontmatter@RRAPformat}
{}{}
}%
\makeatother

\begin{document}

\preprint{AIP/123-QED}

\title[Quantum Algorithm Framework for Phase-Contrast Transmission Electron Microscopy Image Simulation]{Quantum Algorithm Framework for Phase-Contrast Transmission Electron Microscopy Image Simulation}

\author{Sean D. Lam}
\affiliation{Department of Physics, Colorado College, Colorado Springs, CO, 80903, USA}
\affiliation{Department of Chemistry \& Biochemistry, Colorado College, Colorado Springs, CO, 80903, USA}
\author{Roberto dos Reis}
\email{roberto.reis@northwestern.edu}
\affiliation{Department of Materials Science \& Engineering, Northwestern University, Evanston, IL, 60208, USA}
\affiliation{The NUANCE Center, Northwestern University, Evanston, IL, 60208, USA}

\date{\today}

\begin{abstract}
We present a quantum algorithmic framework for simulating phase-contrast transmission electron microscopy (CTEM) image formation using a fault-tolerant, gate-based quantum circuit model. The electron wavefield on an $N\times N$ grid is amplitude-encoded into a $2\log_2 N$-qubit register. Free-space propagation and objective-lens aberrations are implemented via two-dimensional quantum Fourier transforms (QFTs) and diagonal phase operators in reciprocal space, while specimen interaction is modeled under the phase-object approximation (POA) in a single-slice multislice formulation as a position-dependent phase grating. We validate projected potentials, contrast transfer function (CTF) behavior, and image contrast trends against classical multislice simulations for MoS$_2$ over experimentally relevant parameters, and provide resource estimates and key assumptions that determine end-to-end runtime. \rev{Extracting a complete $N\times N$ intensity image by direct sampling requires $O(N^2/\epsilon^2)$ repetitions, which rules out a generic polylogarithmic-cost image-generation algorithm. We therefore analyze end-to-end cost, including state preparation, repeated execution, measurement precision, and the best classical calculation of the same output, and identify the conditions under which limited-output tasks could be candidates for quantum advantage; the present benchmarks establish implementation agreement and do not demonstrate such an advantage.} This framework provides a physics-grounded mapping from CTEM theory to quantum circuits and establishes a baseline for extending toward full multislice and inelastic scattering models.
\end{abstract}

\maketitle

\begin{quotation}
Transmission electron microscopy (TEM) forms images by recording the interference of electron waves that have acquired specimen- and lens-dependent phase shifts, and conventional phase-contrast TEM (CTEM) is routinely modeled using phase-object approximations (POA) and contrast transfer function (CTF) theory.\cite{Kirkland,Reimer} Classical multislice simulations~\cite{cowmood,krivanek} provide quantitative CTEM image formation models \rev{whose cost grows with grid size, specimen thickness, and the number of structural configurations considered, which can make large fields of view and extensive structural ensembles computationally demanding.} In this work, we recast CTEM image formation as a quantum circuit, amplitude-encoding the electron wavefield on an $N\times N$ grid into a register of qubits and implementing free-space propagation and objective-lens aberrations via quantum Fourier transforms and diagonal phase operators, thereby establishing a physics-grounded framework that connects established CTEM theory to emerging fault-tolerant quantum hardware.
\end{quotation}

\section{\label{sec:intro}Introduction}

\begin{figure}[t]
\centering
\includegraphics[width=\linewidth]{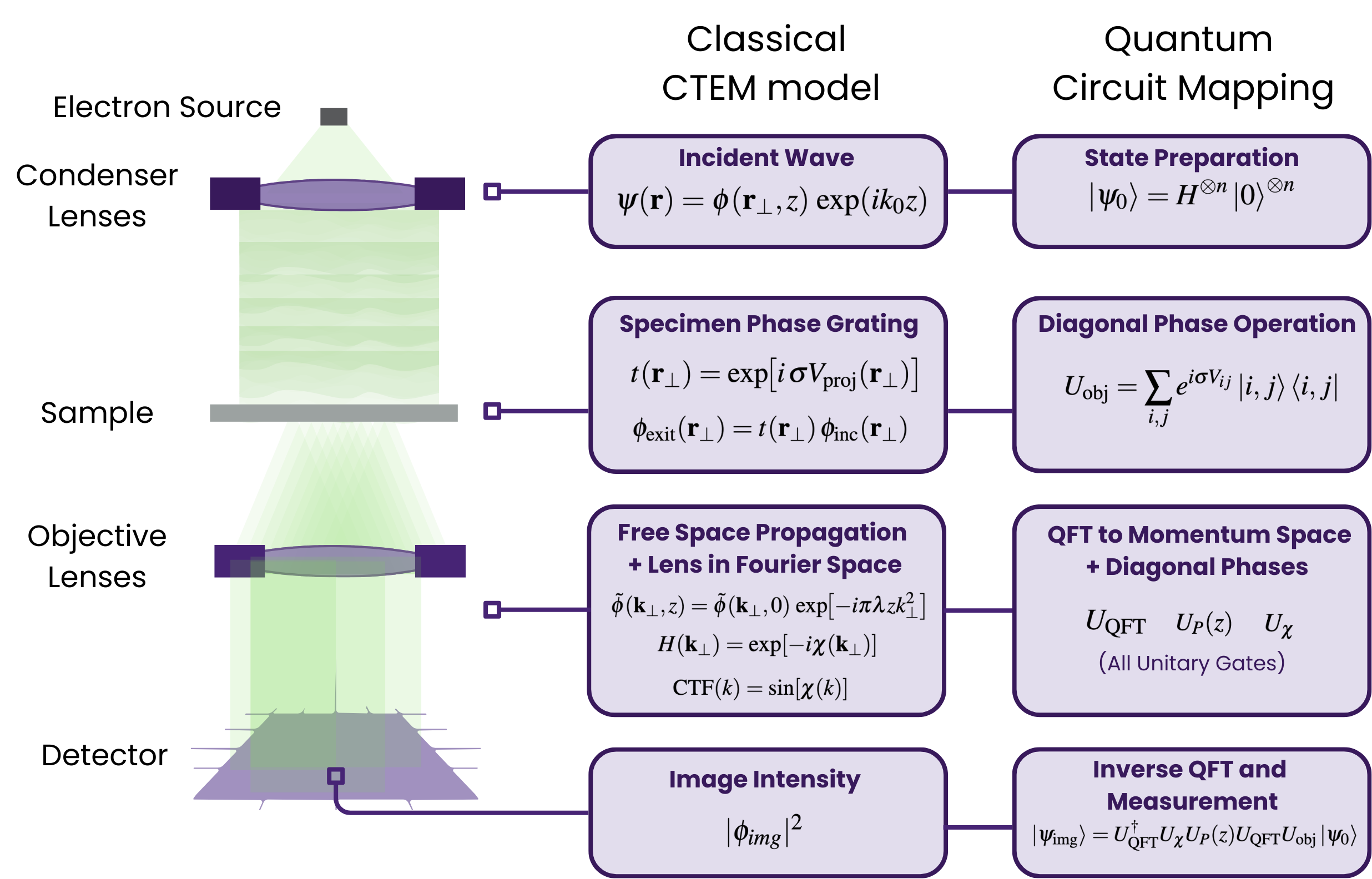}
\caption{Simplified CTEM column schematic (left) and mapping between the classical CTEM image-formation model and the quantum circuit representation (right). The specimen acts as a phase grating for a phase object, free-space propagation and lens aberrations are implemented via Fourier-space phase factors or QFT-based diagonal unitaries, and the recorded intensity corresponds to $\lvert \phi_{\mathrm{img}} \rvert^2$ or $\lvert \psi_{ij} \rvert^2$ at the detector.}
\label{fig:ctem_schematic}
\end{figure}

Transmission electron microscopy (TEM) routinely achieves sub-\AA{}ngstrom resolution by exploiting the short de Broglie wavelength of high-energy electrons to image atomic structures, defects, and interfaces in materials.\cite{haider1998electron, OAM_SubAngstrom_2001, Dahmen_TEAM_2009, bell2010sub} In conventional transmission electron microscopy (CTEM), image contrast arises from the coherent interference of an electron wave that has acquired a specimen-dependent phase modulation and the aberration-induced phase shifts of the objective lens.\cite{Reimer, williams2009transmission} For primarily phase objects, this process is well described by phase-object approximations (POA) combined with free-space propagation and contrast transfer function (CTF) theory~\cite{lentzen2008contrast}, and is routinely simulated using the multislice algorithm.\cite{cowmood,Kirkland} Figure~\ref{fig:ctem_schematic} summarizes the CTEM column and the corresponding classical and quantum descriptions of each imaging stage.

Accurate simulation is now a core component of quantitative electron microscopy, underpinning the interpretation of high-resolution phase contrast, defect imaging, multidimensional STEM modalities and automated analysis workflows.\cite{Kilaas1997,Ophus2019_4DSTEMReview,ML_STEM_2022} A conventional multislice calculation on an $N\times N$ grid requires storing $O(N^2)$ complex amplitudes per slice and performing $O(N^2\log N)$ operations per Fast Fourier Transform (FFT).\cite{PRISM_fastSTEM,abtem} \rev{The practical cost depends on grid size, specimen thickness, potential construction, precision, implementation, and available CPU/GPU hardware; large grids, thick or structurally complex samples, extensive structural ensembles, and real-time integration with experiments can be computationally demanding.}

Quantum computing offers a fundamentally different route to simulating electron wave propagation by representing the discretized wavefunction as a quantum state and implementing propagation and lens action as unitary operators.\cite{nielsen,Reis} Recent work has proposed quantum algorithms for the multislice method that use quantum Fourier transforms (QFTs) to accelerate Fresnel propagation and circuit constructions to encode the specimen-induced phase shifts.\cite{wang-original,wang-improved} While these studies demonstrate that individual multislice propagation steps can be implemented with polylogarithmic gate depth, their analyses assume abstract oracles (black-box operators that apply specimen-dependent phases without specifying their internal arithmetic implementation) for specimen phase operators without explicit arithmetic circuit constructions or validation against realistic atomic potentials. Moreover, the question of end-to-end quantum advantage (regimes where a quantum algorithm achieves asymptotically lower computational cost than any known classical algorithm for the same task), accounting for state preparation, measurement complexity, and synthesis of specimen-dependent phase operators from \textit{ab initio} calculations, remains open.

At the same time, advances in quantum hardware and quantum electron optics are bringing quantum-enhanced microscopy closer to practice. Quantum logic gates acting on free electrons in a TEM column have been demonstrated,\cite{loffler2023quantum} while quantum metrology concepts such as multi-pass TEM and quantum electron microscopy architectures promise substantial gains in information per dose for beam-sensitive samples.\cite{juffmann2017multi,kruit2016designs}

However, the question of whether quantum algorithms can accelerate classical CTEM image simulation remains open. Recent quantum multislice proposals\cite{wang-original,wang-improved} have demonstrated that individual propagation steps can be implemented with polylogarithmic gate depth, but a complete analysis must account for state preparation, measurement complexity, and the cost of synthesizing specimen-dependent phase operators from realistic atomic potentials. Moreover, the primary output of CTEM simulation, a full $N\times N$ intensity image, requires measuring all pixel values, which generically scales with the image size regardless of the internal circuit depth.

In this work, we establish and validate a complete quantum algorithmic framework for CTEM image formation that provides:
\begin{enumerate}
\item A physics-grounded mapping from phase-object-approximation (POA) CTEM theory to fault-tolerant quantum circuits, with explicit arithmetic circuit constructions for specimen potentials encoded from Kirkland/PAW parameterizations (Appendix~\ref{app:diagonal_synthesis}), extending beyond prior oracle-based proposals\cite{wang-original,wang-improved} to gate-level synthesis validated against classical codes;
\item Quantitative validation of projected potentials, contrast transfer functions, and image formation against classical multislice simulations for MoS$_2$ across experimentally relevant parameters, achieving exact numerical agreement (correlation $\rho = 1.000000$, MSE $\sim 10^{-24}$) at floating-point precision (Appendix~\ref{app:validation});
\item Detailed resource estimates including gate counts, ancilla requirements, and measurement complexity for both near-term NISQ demonstrations (Appendix~\ref{app:ibm_hardware}) and fault-tolerant implementations (Table~\ref{tab:resources}, Appendix~\ref{app:prototype_ft}), with explicit analysis of the $O(N^2/\epsilon^2)$ measurement bottleneck for full-image reconstruction;
\item \rev{An illustrative ancilla-assisted construction showing how a phase-sensitive scalar observable can be estimated directly from the amplitude-encoded wavefunction (Sec.~\ref{sec:phase_sensitive_observables}, Appendix~\ref{app:phase_disc}); the same quantity is also computable from a classical simulated wavefield, and no advantage is claimed for it;}
\item \rev{A consolidated end-to-end cost analysis (Sec.~\ref{sec:advantage}) that separates per-run circuit cost, the number of repetitions required for a given observable and precision, the classical output size, and the best classical calculation of the same output.}
\end{enumerate}

\rev{The present benchmarks establish implementation agreement; they do not demonstrate a practical computational advantage. Section~\ref{sec:advantage} states the conditions under which limited-output tasks could become candidates for one.} This framework establishes a baseline for extending toward full multislice, inelastic scattering, and many-body correlation effects.

\subsection{Computational challenges in CTEM simulation}

Classical CTEM simulations are typically performed with the multislice algorithm~\cite{cowmood}, which alternates specimen-induced phase gratings and free-space Fresnel propagation steps implemented via FFTs.\cite{Kirkland,PRISM_fastSTEM} For an $N\times N$ real-space grid, each slice requires storing $O(N^2)$ complex amplitudes and performing at least two FFTs with $O(N^2\log N)$ arithmetic operations per FFT. Even with GPU-accelerated implementations and algorithmic improvements such as PRISM~\cite{PRISM_fastSTEM} or lattice multislice~\cite{LatticeMultislice} methods, the overall cost scales linearly with the number of slices and remains dominated by repeated FFTs over large grids\rev{, growing as field of view, resolution, and thickness increase}. Each slice requires two 2D FFTs (forward and inverse transformations for convolution in real space); for $S$ slices, the total runtime scales as $O(S\,N^2\log N)$ with a memory footprint of $O(N^2)$.\cite{hosokawa2015benchmark}

These scaling laws already limit routine use of full-fidelity multislice simulations for modern high-resolution TEM and scanning TEM (STEM) datasets. For example, a $2048\times 2048$ image sampled at sub-\AA{}ngstrom resolution requires $\sim 4.2\times 10^6$ complex amplitudes per slice and on the order of $10^8$ arithmetic operations per FFT; a typical thickness of tens to hundreds of slices can therefore lead to compute times \rev{ranging from seconds to hours per structural configuration, depending on implementation and hardware}.\cite{PRISM_fastSTEM,abtem,van2015fdes,barthel2012time,grillo2013stem_cell, allen2015modelling,hosokawa2015benchmark} \rev{Systematic exploration of specimen configurations (compositions, defect structures, orientations, thickness) multiplies this cost by the number of configurations, whereas optical parameters can be swept cheaply from a cached exit wave (Sec.~\ref{sec:classical_comparison}); real-time integration of CTEM simulations into experimental feedback loops can remain challenging without substantial computing resources.}

\subsection{Quantum encoding of CTEM wavefunctions}

The central observation enabling a quantum approach to CTEM simulation is that the discretized electron wavefunction is a complex-valued field and can be represented as a pure quantum state on a register of qubits.\cite{nielsen,wang-original,wang-improved} An $N\times N$ grid of complex amplitudes can be mapped to a register of $2\log_2 N$ qubits via two-dimensional amplitude encoding (a mapping in which the entries of a classical data vector are stored as the amplitudes of a quantum state, requiring only $O(\log N)$ qubits for an $N$-element vector), so that each real-space coordinate $(x_i,y_j)$ corresponds to a computational basis state and its complex wave amplitude appears as the associated probability amplitude.\cite{wang-improved} In this representation, the operations central to CTEM, free-space propagation and objective-lens aberrations, can be expressed as unitary transformations generated by quadratic (Fresnel) and higher-order (aberration) phase factors in real or reciprocal space.\cite{wang-original,wang-improved}

On a fault-tolerant (or error-mitigated) quantum processor, these unitaries can be implemented with gate complexity polynomial in the number of qubits by exploiting the quantum Fourier transform (QFT) to move between real-space and momentum-space bases.\cite{nielsen} Since an $N\times N$ wavefield is encoded in $n=2\log_2 N$ qubits, the QFT-based propagation primitives scale as $\mathrm{poly}(n)=\mathrm{poly}(\log N)$ in circuit depth, in contrast to the $O(N^2\log N)$ arithmetic cost of classical FFT-based propagation on an explicit grid.\cite{nielsen} Amplitude encoding also compresses the explicit $O(N^2)$ storage of complex amplitudes to $O(\log N)$ qubits, with additional overheads determined by state preparation and measurement requirements.\cite{nielsen} \rev{As in a classical multislice calculation, the full complex wavefunction $\psi = A e^{i\phi}$ is retained throughout the computation; the quantum representation differs in how quantities derived from it must be extracted, namely by repeated measurement.}\cite{UniversalQEM}

While amplitude encoding compresses the $O(N^2)$ classical memory to $O(\log N)$ qubits and QFT-based propagation achieves polylogarithmic circuit depth, extracting a complete $N\times N$ intensity image via repeated projective measurement generally requires $O(N^2/\epsilon^2)$ shots to resolve all pixel values with relative error $\epsilon$~\rev{ (for pixels near the mean probability $1/N^2$; fainter pixels require more)}.\cite{schuld} This measurement overhead is unavoidable for tasks that demand the full classical image as output and represents the dominant cost for large grids, \rev{outweighing the favorable per-run circuit scaling} for full-image reconstruction. Direct comparison to state-of-the-art GPU-accelerated classical codes (\textsc{abTEM},\cite{abtem} PRISM\cite{PRISM_fastSTEM}) producing $2048\times 2048$ images in seconds indicates that quantum advantage for this baseline task is unlikely in the near term.

\rev{Any potential quantum advantage is therefore task dependent. It is restricted to suitably chosen tasks requiring limited classical output, such as selected Fourier-space probabilities or bounded expectation values, and only if their state-preparation, repetition, and measurement costs remain favorable relative to the best classical algorithm for the same output. Sparse output is a necessary design consideration, not sufficient evidence of advantage. Extensions beyond the POA (inelastic scattering, dynamical diffraction) may change this balance, but whether they do is an open question. We analyze these conditions in Sec.~\ref{sec:advantage}.}

The remainder of this paper is organized as follows. Section~\ref{sec:methods} presents the theoretical foundation of CTEM propagation and the quantum circuit implementation, including specimen potential encoding, contrast transfer function realization, and resource scaling analysis. Section~\ref{sec:results} validates the quantum algorithm against classical multislice simulations for a MoS$_2$ benchmark across a comprehensive microscope parameter space. Section~\ref{sec:discussion} \rev{analyzes end-to-end cost and the conditions for potential quantum advantage}, and discusses extensions to realistic experimental conditions. Section~\ref{sec:conclusion} summarizes the main findings and outlines future directions.

\section{Methods}
\label{sec:methods}

We implement quantum CTEM image formation by mapping a single-slice multislice realization of the phase-object approximation (POA) to a fault-tolerant quantum circuit that encodes the electron wavefunction as a superposition over an $N\times N$ spatial grid, with propagation and lens aberrations realized via quantum Fourier transforms and diagonal phase operators. This section presents: (i)~the classical POA framework and paraxial propagation theory (Sec.~\ref{sec:wpoa}), (ii)~the amplitude encoding and quantum gate construction (Sec.~\ref{sec:circuit}), (iii)~the atomic potential parameterization used for the MoS$_2$ benchmark (Sec.~\ref{sec:potential_method}), (iv)~the contrast transfer function implementation including partial coherence considerations (Sec.~\ref{sec:ctf_method}), and (v)~detailed resource estimates for fault-tolerant operation (Sec.~\ref{sec:resources}).

\subsubsection{Electron wave propagation in CTEM}
\label{sec:wpoa}

High-energy electrons in a TEM are well described by a scalar wavefunction $\psi(\mathbf{r})$ obeying a Schr\"odinger-type equation with an effective electrostatic potential $V(\mathbf{r})$.\cite{Kirkland,Reimer, egerton2005physical} Under the usual paraxial approximation, the rapidly oscillating carrier along the optical axis $z$ can be factored out as
\begin{equation}
\psi(\mathbf{r}) = \phi(\mathbf{r}_\perp,z)\,\exp(i k_0 z),
\end{equation}
where $\mathbf{r}_\perp=(x,y)$ is the transverse coordinate, $k_0=2\pi/\lambda$, and $\phi$ denotes the slowly varying transverse envelope.\cite{Reimer}

For sufficiently thin specimens and weak scattering, the phase-object approximation (POA) models the specimen transmission function as a pure phase modulation acting on an incident plane wave.\cite{Reimer, egerton2005physical} We adopt the POA, in which the specimen transmission function retains the full phase grating $\exp[i\sigma V_{\mathrm{proj}}(\mathbf{r}_\perp)]$ rather than its first-order linearization $1 + i\sigma V_{\mathrm{proj}}(\mathbf{r}_\perp)$ (the weak phase object approximation, WPOA). The latter is recovered as a first-order expansion of the POA and is valid only when $\sigma V_{\mathrm{proj}} \ll 1$; we use it below (Sec.~\ref{sec:phase_sensitive_observables}) only for comparison in the phase-sign ambiguity discussion.

The projected potential $V_{\mathrm{proj}}(\mathbf{r}_\perp)=\int V(\mathbf{r}_\perp,z)\,dz$ then defines a transmission function
\begin{equation}
t(\mathbf{r}_\perp) = \exp\!\bigl[i\,\sigma V_{\mathrm{proj}}(\mathbf{r}_\perp)\bigr],
\end{equation}
with the interaction constant $\sigma$ set by the electron wavelength and relativistic mass.\cite{egerton2005physical} The exit wave immediately after the specimen is
\begin{equation}
\phi_{\mathrm{exit}}(\mathbf{r}_\perp) = t(\mathbf{r}_\perp)\,\phi_{\mathrm{inc}}(\mathbf{r}_\perp).
\end{equation}

Propagation from the specimen to the image (or back focal) plane is described by a paraxial free-space propagator generated by the transverse Laplacian,
\begin{equation}
P(z) = \exp\!\bigl(-i\pi\lambda z \nabla_\perp^2\bigr).
\end{equation}
In reciprocal space, this becomes a simple $\mathbf{k}_\perp$-dependent phase factor,
\begin{equation}
\tilde{\phi}(\mathbf{k}_\perp,z) = \tilde{\phi}(\mathbf{k}_\perp,0)\,
\exp\!\bigl[-i\pi\lambda z k_\perp^2\bigr],
\end{equation}
where $\mathbf{k}_\perp$ is the transverse spatial frequency.\cite{Reimer}

Objective-lens aberrations introduce an additional phase shift $\chi(\mathbf{k}_\perp)$, so that the total transfer function in Fourier space is
\begin{equation}
H(\mathbf{k}_\perp) = \exp\!\bigl[-i\chi(\mathbf{k}_\perp)\bigr],
\end{equation}
and, for an axially symmetric lens,
\begin{equation}
\chi(k) = \pi\lambda\Delta f\,k^2 + \tfrac{1}{2}\pi\lambda^3 C_3 k^4 + \tfrac{1}{3}\pi\lambda^5 C_5 k^6 + \cdots,
\label{eq:chi_function}
\end{equation}
where $\Delta f$ is the defocus and $C_3, C_5$ are spherical aberration coefficients (we use $C_3$ consistently throughout).\cite{rose,elferich2024ctffind5} For phase-contrast CTEM, the resulting contrast transfer function (CTF) is
\begin{equation}
\mathrm{CTF}(k) = \sin\!\bigl[\chi(k)\bigr],
\label{eq:ctf_equation}
\end{equation}
which determines the spatial frequencies (and associated phases) transferred from the exit wave to the recorded image intensity.\cite{Reimer}

\paragraph{Resolution definitions.}
\begin{revblock}
We distinguish the first positive CTF zero $k_0$, expressed in \AA$^{-1}$, from its reciprocal length $d_0=1/k_0$, expressed in \AA. These quantities depend on the selected defocus and aberrations. The term Scherzer defocus is reserved for the optimum of the specified third-order spherical-aberration model, $\Delta f_{\mathrm{Sch}}=-\sqrt{4C_3\lambda/3}$ under the convention of Eq.~\eqref{eq:chi_function}; at this defocus the phase $\chi(k)$ reaches a minimum of $-2\pi/3$ and returns to zero at $k_0=\sqrt{2\lvert\Delta f_{\mathrm{Sch}}\rvert/(\lambda^2 C_3)}$.

For the hypothetical lens with $C_3=0$ and all higher-order aberrations neglected, the nonzero defocus used in Fig.~\ref{fig:ctf_cs_comparison}(a) is a selected illustrative setting, not Scherzer defocus; its first zero, where $\chi=-\pi$, lies at $k_0=(\lambda\lvert\Delta f\rvert)^{-1/2}$. Practical information transfer additionally depends on partial coherence, apertures, detector response, and the criterion used to define an information limit; the coherent quantities $k_0$ and $d_0$ are not estimates of it.
\end{revblock}

\subsubsection{Quantum algorithm design}
\label{sec:circuit}

We represent the discretized CTEM wavefield on an $N\times N$ real-space grid as an amplitude-encoded quantum state on $n = 2\log_2 N$ qubits,\cite{wang-improved,Ashhab2022,gonzalez2024efficient}
\begin{equation}
\ket{\psi} = \sum_{i,j=0}^{N-1} \psi_{ij}\,\ket{i}\ket{j}
= \sum_{i,j=0}^{N-1} \psi_{ij}\,\ket{i,j},
\label{eq:amplitude_encoding}
\end{equation}
where $\ket{i}$ and $\ket{j}$ denote computational basis states for the $x$ and $y$ coordinates, respectively. We use both the factored notation $\ket{i}\ket{j}$ and the compact notation $\ket{i,j}$ interchangeably throughout; they represent the same tensor product state $\ket{i}_x \otimes \ket{j}_y$ in the $n$-qubit Hilbert space.

The specimen interaction under the POA is implemented as a position-dependent phase operator. Given a projected potential sampled on the discrete grid, $V_{ij}\approx V_{\mathrm{proj}}(x_i,y_j)$, the transmission function $t_{ij} = \exp\!\bigl(i\sigma V_{ij}\bigr)$ is realized by the diagonal unitary
\begin{equation}
U_{\mathrm{obj}} = \sum_{i,j} e^{i\sigma V_{ij}}\ket{i,j}\bra{i,j},
\end{equation}
which can be decomposed into a sequence of controlled $R_z$ rotations conditioned on the binary representation of $(i,j)$ (or, more generally, synthesized via reversible evaluation of $V_{ij}$ into a work register followed by phase kickback).\cite{wang-improved} Acting on $\ket{\psi_0}$, this yields the exit wave $\ket{\psi_{\mathrm{exit}}}=U_{\mathrm{obj}}\ket{\psi_0}$.

Free-space propagation and lens aberrations are implemented in momentum space using the quantum Fourier transform (QFT). A two-dimensional QFT, $U_{\mathrm{QFT}} = \mathrm{QFT}_x\otimes\mathrm{QFT}_y$, maps position basis states $\ket{i,j}$ to transverse spatial-frequency states $\ket{k_x,k_y}$ with gate complexity $O(n^2)$.\cite{nielsen} In this basis, paraxial propagation over a distance $z$ corresponds to a $k$-dependent phase,
\begin{equation}
U_{P}(z) = \sum_{k_x,k_y}
\exp\!\bigl[-i\pi\lambda z (k_x^2+k_y^2)\bigr]
\ket{k_x,k_y}\bra{k_x,k_y},
\end{equation}
while the objective lens is represented by
\begin{equation}
U_{\chi} = \sum_{k_x,k_y}
\exp\!\bigl[-i\chi(k)\bigr]
\ket{k_x,k_y}\bra{k_x,k_y},
\end{equation}
with $\chi(k)$ given by Eq.~\ref{eq:chi_function}. Both operators are diagonal in the momentum basis and can be synthesized from single-qubit and controlled-phase gates acting on the qubit registers encoding $k_x$ and $k_y$.\cite{wang-original,wang-improved}

\begin{figure}[H]
\centering
\includegraphics[width=\linewidth]{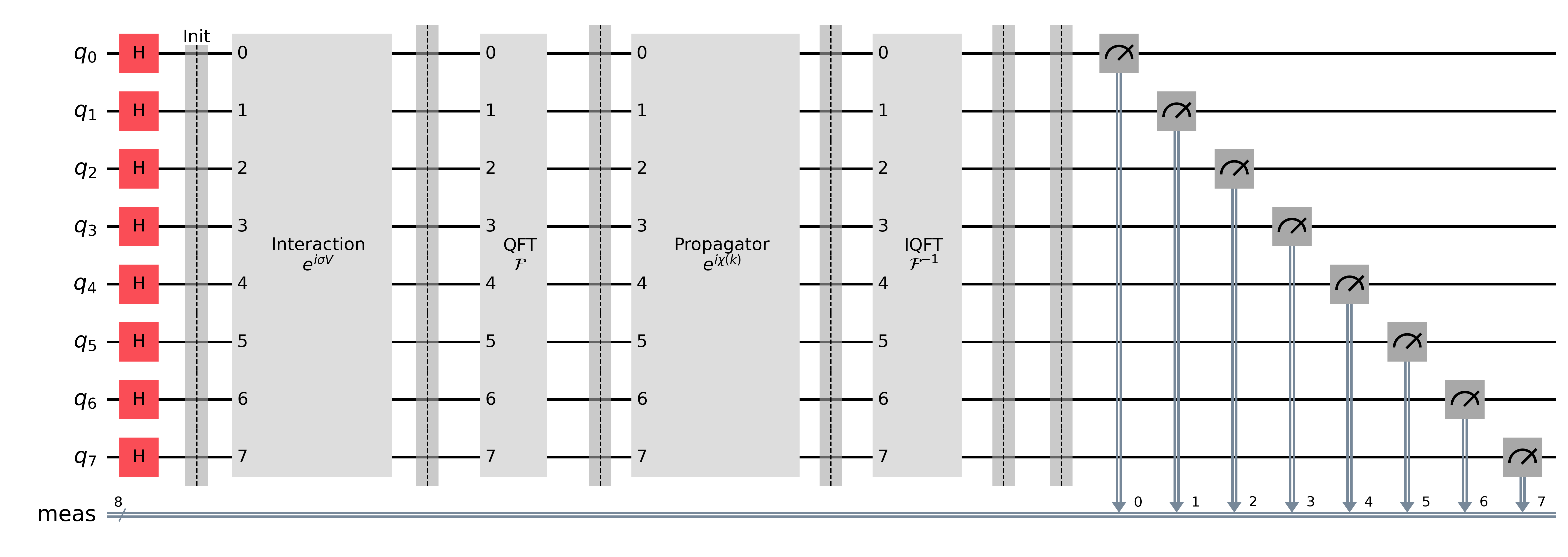}
\caption{Example of quantum CTEM circuit implementing state preparation, specimen phase grating $U_{\mathrm{obj}}$, two-dimensional QFT $U_{\mathrm{QFT}}$, Fresnel propagator $U_P$, lens phase $U_{\chi}$, and inverse QFT, followed by measurement of the image-plane intensity. The schematic is shown here for a $16\times 16$ grid ($n=8$ qubits).}
\label{fig:ctem_circuit}
\end{figure}

The full CTEM imaging sequence is therefore realized by the circuit
\begin{equation}
\ket{\psi_{\mathrm{img}}}
= U_{\mathrm{QFT}}^\dagger U_{\chi} U_{P}(z) U_{\mathrm{QFT}} U_{\mathrm{obj}} \ket{\psi_0},
\label{eq:ctem_equation}
\end{equation}
which maps the initial plane wave to the image-plane wavefunction. A schematic of the corresponding quantum circuit is shown in Figure~\ref{fig:ctem_circuit}.

The quantum circuits described above are implemented in the QuScope package, an open-source Python library for quantum simulation of electron microscopy.\cite{quscope_github} QuScope provides state preparation, circuit construction, Qiskit-based simulation, and classical reference comparison utilities: it implements the CTEM imaging sequence [Eq.~\ref{eq:ctem_equation}] as a sequence of Qiskit circuit operations---state preparation via \texttt{initialize()}, diagonal phase operators via controlled-$R_z$ decompositions, and QFT via the Qiskit \texttt{QFT()} circuit block (Qiskit v2.2.1)---and is available at \url{https://github.com/QuScope/QuScope}. Classical reference simulations throughout this work use \textsc{abTEM} v1.0.5.\cite{abtem}

\subsubsection{Sample potential and projected phase}
\label{sec:potential_method}

To benchmark the quantum algorithm against established multislice simulations, we construct projected electrostatic potentials using the Kirkland tabulated parameterization of atomic scattering factors.\cite{Kirkland} For each atomic species, the projected potential is approximated as a sum of Gaussians,
\begin{equation}
V_{\mathrm{atom}}(r) = \sum_{m=1}^{M} a_m \exp\!\left(-\pi r^2/b_m\right),
\end{equation}
with tabulated coefficients $\{a_m,b_m\}$ chosen to reproduce electron scattering factors over the relevant spatial-frequency range.\cite{Kirkland}
The projected potential $V_{\mathrm{proj}}(\mathbf{r}_\perp)$ for a given structure is then obtained by superposing the atomic contributions on a discrete real-space grid, followed by periodic boundary conditions and vacuum padding to suppress wrap-around artifacts and spurious interactions.\cite{Kirkland,abtem}

For the MoS$_2$ test system studied here, we generate supercells by tiling the hexagonal unit cell ($a\approx 3.16$~\AA{}) to fill an $N\times N$ simulation box and compare the resulting $V_{\mathrm{proj}}$ to the output of the \textsc{abTEM} package.\cite{abtem} Both the quantum implementation and the \textsc{abTEM}\cite{abtem} reference simulations use the identical parameterization---the same tabulated Gaussian coefficients $\{a_m, b_m\}$ and the same real-space sampling grid---ensuring that any differences in the resulting CTEM images reflect the propagation and imaging stages rather than the underlying potential model.
No element-specific scaling factor is needed to align the two potentials, and the spatial distribution and column intensities of $V_{\mathrm{proj}}$ agree with \textsc{abTEM} to floating-point precision.

\subsubsection{CTF implementation and parameter space}
\label{sec:ctf_method}

Objective-lens aberrations are incorporated in the quantum algorithm through the phase function $\chi(k)$ in Eq.~\ref{eq:chi_function}, encoded as a diagonal unitary $U_{\chi}$ in the momentum basis. For the present study, we restrict to axially symmetric aberrations and vary defocus $\Delta f$ and third-order spherical aberration $C_3$ over ranges representative of modern 80--300~kV microscopes.\cite{Kirkland} The corresponding phase shifts are discretized on the same $\mathbf{k}_\perp$ grid used for propagation and implemented via controlled-phase gates acting on the qubit registers that encode $k_x$ and $k_y$.\cite{nielsen,wang-improved}

\begin{figure*}[!t]
\centering
\includegraphics[width=1\linewidth]{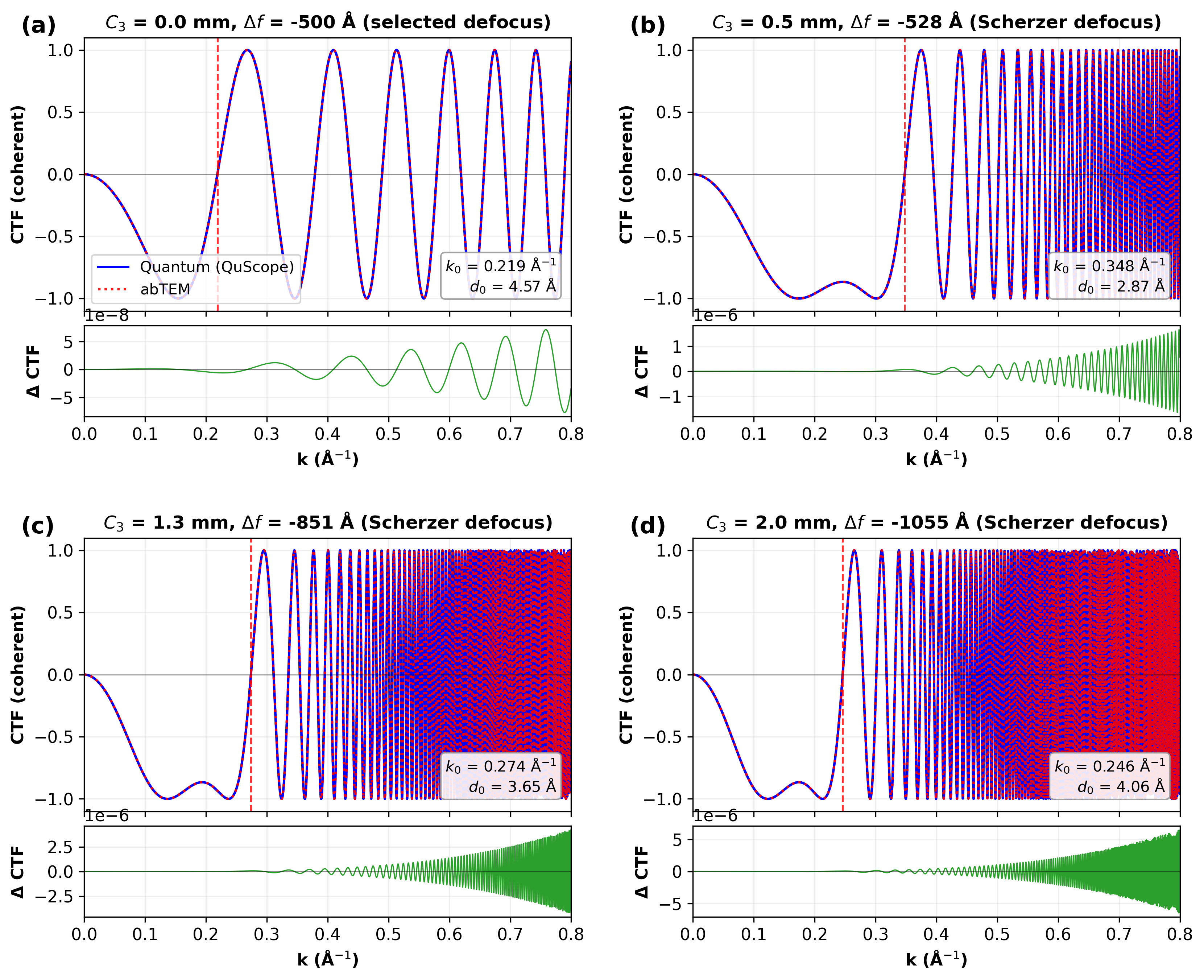}
\caption{\rev{Coherent} quantum-computed contrast transfer function for varying spherical aberration at 80~kV in comparison to \textsc{abTEM}. \rev{(a)~Hypothetical aberration-free lens, $C_3 = 0$~mm, at a selected defocus $\Delta f = -500$~\AA{} (not a Scherzer condition); (b)--(d)~$C_3 = 0.5$, $1.3$, and $2.0$~mm at the Scherzer defocus $\Delta f_{\mathrm{Sch}}=-\sqrt{4C_3\lambda/3} = -528$, $-851$, and $-1055$~\AA, respectively. The vertical dashed line in each panel marks that panel's first positive CTF zero $k_0$ (excluding $k=0$), and the annotation gives $k_0$ and $d_0=1/k_0$. No spatial- or temporal-coherence envelopes are applied. The residual panel beneath each curve shows $\Delta\mathrm{CTF}(k) = \mathrm{CTF}_{\mathrm{quantum}}(k) - \mathrm{CTF}_{\mathrm{abTEM}}(k)$, with both codes evaluated on the same $k$ grid; it remains below $10^{-5}$ throughout and is consistent with the $\sim\!4\times10^{-9}$ relative difference between the electron wavelengths computed by the two codes, amplified by the large accumulated phase $\chi(k)$ at high $k$. The display is limited to $k\le0.8$~\AA$^{-1}$, below the Nyquist frequency of the $0.5$~\AA{} sampling and in a range where $\chi(k)$ is adequately sampled.}
}
\label{fig:ctf_cs_comparison}
\end{figure*}

To validate the correctness of the CTF implementation, we compute the phase-contrast transfer function [Eq.~\ref{eq:ctf_equation}] for a set of accelerating voltages between 80 and 300~kV, defocus values spanning typical underfocus and overfocus conditions, and $C_3$ values ranging from aberration-corrected ($C_3\approx 0$) to conventional uncorrected instruments. Figure~\ref{fig:ctf_cs_comparison} shows the computed \rev{coherent} CTF for four representative spherical aberration values at 80~kV\rev{, evaluated on the same spatial-frequency grid as an independent \textsc{abTEM} calculation. The first positive zeros extracted from the quantum calculation, $k_0=0.219$, $0.348$, $0.274$, and $0.246$~\AA$^{-1}$ ($d_0=4.57$, $2.87$, $3.65$, and $4.06$~\AA) for $C_3=0$, $0.5$, $1.3$, and $2.0$~mm, agree with the analytical expressions above to within the grid spacing ($\Delta k\approx2.4\times10^{-4}$~\AA$^{-1}$), and the pointwise difference from \textsc{abTEM} is below $10^{-5}$ over the displayed band}, providing a stringent test of the gate-level realization of microscope optics. A complete through-focus series at fixed $C_3=1.3$~mm is provided in Appendix Figure~\ref{app:ctf_defocus} and a comprehensive voltage-defocus parameter grid (Appendix Figure~\ref{app:ctf_voltage}) demonstrate systematic agreement with analytical CTF theory across a variety of experimental operating range.

The idealized CTF in Eq.~\eqref{eq:ctf_equation} assumes perfect temporal and spatial coherence. In practice, \rev{finite illumination convergence and energy spread} introduce envelope functions that damp high spatial frequencies:
\begin{equation}
\label{eq:ctf_eff}
\text{CTF}_{\text{eff}}(k) = E_s(k)\, E_c(k)\, \sin\bigl[\chi(k)\bigr],
\end{equation}
\begin{revblock}
where, in the quasi-coherent approximation and for the convention of Eq.~\eqref{eq:chi_function}, the spatial-coherence envelope is
\begin{equation}
E_s(k) = \exp\!\left[-\pi^2\alpha^2\left(\Delta f\,k + \lambda^2 C_3 k^3\right)^2\right],
\label{eq:spatial_envelope}
\end{equation}
with $\alpha$ the $1/e$ semi-angle of a Gaussian distribution of illumination angles, and the temporal-coherence (chromatic) envelope is
\begin{equation}
E_c(k) = \exp\!\left(-\tfrac{1}{2}\pi^2\lambda^2\delta^2 k^4\right),
\qquad
\delta = C_c\,\frac{\sigma_E}{E_0},
\label{eq:chromatic_envelope}
\end{equation}
with $C_c$ the chromatic aberration coefficient and $\delta$ the standard deviation of the defocus spread produced by an energy spread of standard deviation $\sigma_E$ (other instabilities add in quadrature).\cite{Reimer,Kirkland} If a full-width-at-half-maximum spread $\Delta_{\mathrm{FWHM}}$ is used instead, $\delta^2=\Delta_{\mathrm{FWHM}}^2/(8\ln 2)$. The spatial envelope depends on defocus and $C_3$ through the gradient of $\chi$, so the resulting information limit is condition dependent; we do not evaluate these envelopes in the present work, and no practical information limit is quoted.
\end{revblock}

In the present quantum circuit implementation, we compute the \emph{coherent} CTF (Eq.~\ref{eq:ctf_equation}) to validate the gate-level realization of the lens phase $\chi(k)$ against analytical theory. Incorporating partial coherence and detector response into the quantum framework (either via mixed-state density matrices or ensembles of pure-state runs) is straightforward in principle but increases resource requirements; we discuss this extension in Appendix~\ref{app:coherence}.

\subsubsection{Resource estimates and scaling}
\label{sec:resources}

We consider two implementation regimes: a near-term prototype implementation using direct state-vector synthesis ($O(N^2)$ gate cost, suitable for NISQ devices and small grid sizes) and a fault-tolerant implementation using arithmetic circuit synthesis ($O(\mathrm{poly}(\log N))$ gate depth, requiring error-corrected qubits). The prototype implementation, detailed in Appendix~\ref{app:prototype_ft}, enables the proof-of-concept hardware demonstration reported in Appendix~\ref{app:ibm_hardware}, while the fault-tolerant implementation, analyzed below, \rev{determines the per-run circuit scaling for larger grids; end-to-end cost is analyzed in Sec.~\ref{sec:advantage}}.


\begin{table*}[!t]
\centering
\caption{Fault-tolerant resource estimates for quantum CTEM simulation of a MoS$_2$ $3\times 2$ supercell (18 atoms, 5 Gaussians/atom). \rev{Shot counts are for full-image reconstruction by direct computational-basis sampling, $P=N^2/\epsilon^2$ with $\epsilon=0.01$, which gives a 1\% relative standard error for a pixel at the mean probability $1/N^2$ (fainter pixels require more shots; amplitude estimation is not assumed; Sec.~\ref{sec:advantage}).} See Fig.~\ref{fig:resource_scaling} for graphical scaling trends and Appendix~\ref{app:diagonal_synthesis} for the gate-count breakdown.}
\label{tab:resources}
\begin{tabular}{cccccc}
\hline
Grid & Data & Ancilla & Total & T-gates & Shots \\
& Qubits & (est.) & Logical & (est.) & ($\epsilon=0.01$) \\
\hline
$4\times 4$ & 4 & 30 & 34 & $4.9\times10^5$ & $1.6\times10^5$ \\
$16\times 16$ & 8 & 30 & 38 & $5.3\times10^5$ & $2.6\times10^6$ \\
$64\times 64$ & 12 & 34 & 46 & $5.7\times10^5$ & $4.1\times10^7$ \\
$256\times 256$ & 16 & 42 & 58 & $6.1\times10^5$ & $6.6\times10^8$ \\
$1024\times 1024$ & 20 & 50 & 70 & $6.6\times10^5$ & $1.0\times10^{10}$ \\
\hline
\end{tabular}
\end{table*}

\begin{figure*}[!t]
\centering
\includegraphics[width=\textwidth]{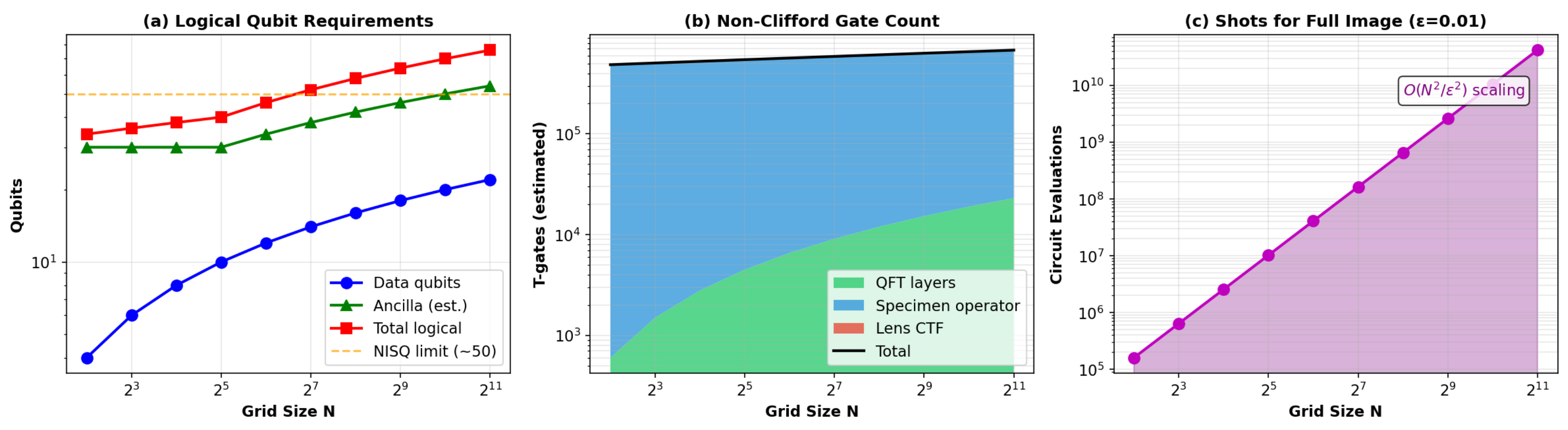}
\caption{Fault-tolerant resource scaling for quantum CTEM simulation of MoS$_2$ (18 atoms, 5 Gaussians/atom). (a)~Logical qubit requirements: data qubits (blue) grow logarithmically as $2\log_2 N$, ancilla (green) remain approximately constant at $\sim 30$--50, and total logical qubits (red) stay below 100 for grids up to $1024\times 1024$. Horizontal dashed line marks a representative NISQ-limit estimate ($\sim 50$ logical qubits). (b)~Non-Clifford (T) gate count breakdown: QFT layers (green), specimen operator $U_{\text{obj}}$ (blue), lens CTF $U_\chi$ (red), and total (black line). Gate counts increase modestly from $\sim 5\times10^5$ to $7\times10^5$ across grid sizes, dominated by $U_{\text{obj}}$ for this fixed atomic structure (90 Gaussian terms). The near-constant behavior confirms polylogarithmic scaling in grid size $N$ for fixed specimen complexity. (c)~Measurement cost: shot count for full $N\times N$ image reconstruction at $\epsilon = 0.01$ relative error scales as $O(N^2/\epsilon^2)$ (purple shaded region), reaching $10^{10}$ circuit evaluations for $1024\times 1024$. This measurement overhead dominates the single-shot circuit cost (panels a--b) by four to six orders of magnitude, representing the primary end-to-end runtime bottleneck for full-image reconstruction tasks and precluding quantum advantage over classical GPU codes for this baseline application.}
\label{fig:resource_scaling}
\end{figure*}

For an $N\times N$ grid encoded in $n = 2\log_2 N$ qubits, the two-dimensional  QFT layers require $O(n^2)$ single- and two-qubit gates with circuit depth $O(n^2)$.\cite{nielsen} The diagonal specimen and lens operators $U_{\text{obj}}$ and $U_\chi$ are synthesized via arithmetic evaluation of the phase functions $\sigma V_{ij}$ and $\chi(k)$, requiring additional ancilla qubits and non-Clifford (T) gates (a Clifford gate---Hadamard, CNOT, $S$, and related gates---can be efficiently simulated classically and does not by itself contribute to quantum advantage, whereas the T gate, $T=\mathrm{diag}(1,e^{i\pi/4})$, is a non-Clifford single-qubit gate that, together with Clifford gates, forms a universal gate set for fault-tolerant quantum computation; the T-gate count is a standard proxy for fault-tolerant resource cost).

\paragraph{Gate-count estimates.}
For the specimen operator, evaluating the projected potential
$V_{ij} = \sum_{\text{atoms}} \sum_{m} a_m \exp(-\pi r^2 / b_m)$ at each grid point involves:
\begin{itemize}
\item Computing squared distances $r^2 = (x_i - x_a)^2 + (y_j - y_a)^2$
      for each atom, requiring $O(n)$ arithmetic operations per atom;
\item Evaluating Gaussian exponentials via polynomial approximation, with cost $\sim 10^3$ to $10^4$ T-gates per term depending on target precision;
\item Accumulating contributions and applying phase kickback
      $R_z(\sigma V_{ij})$ controlled on the position register.
\end{itemize}
For a supercell with $N_{\text{atoms}}$ atoms and $M$ Gaussians per
species, the estimated T-count is $O(N_{\text{atoms}} \times M \times
\text{poly}(n, p))$, where $p$ is the number of bits of precision.
The lens phase $\chi(k)$ similarly involves polynomial evaluation of
$k^2$ and $k^4$ terms, contributing $O(n^2)$ Toffoli gates
($\sim 4$ T-gates each).

Detailed arithmetic circuits and ancilla counts are provided in
Appendix~\ref{app:diagonal_synthesis}. For the $128\times 128$ MoS$_2$
benchmark ($n=14$, 18 atoms, 5 Gaussians/atom), we estimate a total
non-Clifford count of \rev{$\approx 5.9\times10^5$} T-gates per image, excluding
error-correction overhead. \rev{For this fixed specimen, scaling to $2048\times 2048$ ($n=22$) increases
this only to $\approx 6.8\times10^5$ T-gates; the count grows instead with the number of atoms and Gaussian terms (see below). All T-gate figures in this work follow the single cost model stated in Appendix~\ref{app:diagonal_synthesis}, which underlies Table~\ref{tab:resources} and Figs.~\ref{fig:resource_scaling}, \ref{fig:ft_scaling}, and~\ref{fig:prototype_ft_scaling}.}

\paragraph{Logical qubit requirements.}
In addition to the $n$ data qubits, arithmetic evaluation requires
$\sim 30$--$50$ ancilla qubits for fixed-point registers, accumulators,
and scratch space (Table~\ref{tab:resources}).

Fault-tolerant implementation with surface-code error correction at logical
error rate $\sim 10^{-9}$ requires code distance $d \sim 15$--$20$,
corresponding to $\sim 2d^2 \approx 450$--$800$ physical qubits per logical
qubit.\cite{fowler2012surface} For total logical qubit counts of 58
($256\times 256$ grid) and 70 ($1024\times 1024$ grid) from
Table~\ref{tab:resources}, the physical overhead is \rev{$\approx 2.6$--$4.6\times10^4$ and $\approx 3.2$--$5.6\times10^4$ physical qubits,
respectively, excluding magic-state factories}. Non-Clifford (T) gates required for arithmetic evaluation of
diagonal operators are synthesized via magic state distillation from noisy
ancillae.\cite{bravyi2005universal,litinski2019magic}

\paragraph{Measurement complexity.}
Recovering the full $N\times N$ intensity distribution $|\psi_{ij}|^2$ requires
measuring all computational basis amplitudes. \rev{With direct projective sampling, a pixel of probability $p_{ij}=|\psi_{ij}|^2$ is estimated with relative standard error $\epsilon$ after $P\approx(1-p_{ij})/(p_{ij}\epsilon^2)$ shots; for pixels near the mean probability $p_{ij}\approx 1/N^2$, this gives the $O(N^2/\epsilon^2)$ total used in Table~\ref{tab:resources}, while pixels much fainter than the mean require proportionally more. Amplitude estimation, which uses coherent repetitions of the circuit rather than independent shots, does not remove this cost for a full image, because it must be applied separately to each of the $N^2$ outputs (Sec.~\ref{sec:advantage}).} This scaling in image size is unavoidable for tasks that demand the complete classical image and represents the dominant cost for large grids, \rev{outweighing the favorable per-run circuit scaling}.\cite{aaronson2015read, schuld2019quantum}

\paragraph{Scaling regimes for specimen operator cost.}
The T-gate estimates in Table~\ref{tab:resources} exhibit two distinct
scaling behaviors depending on how the problem size $N$ increases:
\begin{enumerate}
\item \textbf{Fixed field-of-view, increasing pixel density:} Holding the atomic structure constant (18 atoms, 90 Gaussian terms for the MoS$_2$ benchmark) while increasing grid resolution from $4\times 4$ to $1024\times 1024$ yields T-counts from $4.9\times10^5$ to $6.6\times10^5$---a modest 35\% increase despite a 256-fold increase in $N$. This near-constant behavior arises because the specimen operator $U_{\text{obj}}$ cost is dominated by evaluating 90 Gaussian functions (\rev{$\sim 5\times10^3$ T-gates each, plus $\sim\!100n$ for the associated distance arithmetic}), with only logarithmic growth \rev{in the arithmetic register width} as $N$ increases. The QFT layers contribute \rev{$\sim 10^3$--$10^4$ T-gates ($9.1\times10^3$ at $128\times128$)} and scale as $O(n^2) = O(\log^2 N)$, \rev{small} compared to $U_{\text{obj}}$. Concretely, in the detailed component breakdown for the $128\times 128$ benchmark (Appendix~\ref{app:diagonal_synthesis}, Table~\ref{tab:detailed_resources}), $U_{\text{obj}}$ alone accounts for \rev{$5.76\times10^5$ of the $5.88\times10^5$ total T-gates---approximately 98\%} of the end-to-end T-count---confirming that specimen-operator synthesis, not QFT-based propagation, dominates the fault-tolerant resource cost for fixed atomic structure. This cost is unavoidable for any circuit that must encode the specimen's projected potential. \rev{A compiled specimen circuit can be reused across optical configurations, but each repetition still executes it, so the specimen operator contributes to every shot (Sec.~\ref{sec:advantage}).}

\item \textbf{Increasing field-of-view (proportional supercell expansion):} If both the number of atoms $N_{\text{atoms}}$ and grid size $N$ increase proportionally (e.g., imaging a $6\times 4$ MoS$_2$ supercell on a $256\times 256$ grid, doubling both dimensions), the T-count scales linearly with $N_{\text{atoms}}$ because each atom contributes $M$ Gaussian evaluations. For example, a $6\times 4$ supercell \rev{(72 atoms, four times the 18 atoms of the $3\times2$ cell) would require $\sim 2.4\times10^6$ T-gates, approximately four times the $3\times 2$ value in Table~\ref{tab:resources}. At fixed sampling, $N_{\text{atoms}}$ grows in proportion to the field-of-view area, i.e., as $N^2$, so in this regime the per-run specimen cost also grows as $N^2$.}
\end{enumerate}

This distinction is crucial for understanding resource scaling in realistic applications: refining images of a fixed structure benefits from polylogarithmic scaling, while simulating larger fields of view (defects, interfaces, grain boundaries) incurs linear overhead in atom count.

\rev{The implications of these per-run costs for end-to-end runtime, once repetitions, output size, and the classical baseline are included, are analyzed in Sec.~\ref{sec:advantage}.}

Figure~\ref{fig:resource_scaling} visualizes the scaling behavior across grid sizes from $4\times 4$ to $1024\times 1024$.~\ref{fig:resource_scaling}(a)
shows the logarithmic growth of data qubits ($2\log_2 N$, blue) versus the approximately constant ancilla count (green), with total logical qubits remaining below 100 even for production-scale grids. Figure~\ref{fig:resource_scaling}(b) illustrates the gate-count breakdown: QFT layers \rev{(green)} scale as $O(n^2)$ while the specimen operator \rev{(blue)} dominates for dense atomic structures, both contributing to a total T-count of $\sim 5\times10^5$ to $7\times10^5$ for the MoS$_2$ benchmark across all tested grid sizes. Figure~\ref{fig:resource_scaling}(c) highlights the measurement bottleneck: shot counts for full-image reconstruction scale quadratically as $O(N^2/\epsilon^2)$ (purple region), reaching $\sim 10^{10}$ circuit evaluations for $1024\times 1024$ at 1\% precision, orders of magnitude larger than the single-shot circuit cost and the dominant factor in end-to-end runtime.

Derivations of the asymptotic scaling  arguments are provided in Appendix~\ref{app:complexity}, and explicit arithmetic circuit constructions for the diagonal operators are detailed
in Appendix~\ref{app:diagonal_synthesis}.

Proof-of-principle hardware execution of the QuScope circuits on IBM quantum processors is detailed in Appendix~\ref{app:ibm_hardware}.

\section{Results and Discussion}
\label{sec:results}

\subsubsection{Projected potential validation}
\label{sec:potential_validation}

\begin{figure*}[!t]
\centering
\includegraphics[width=\textwidth]{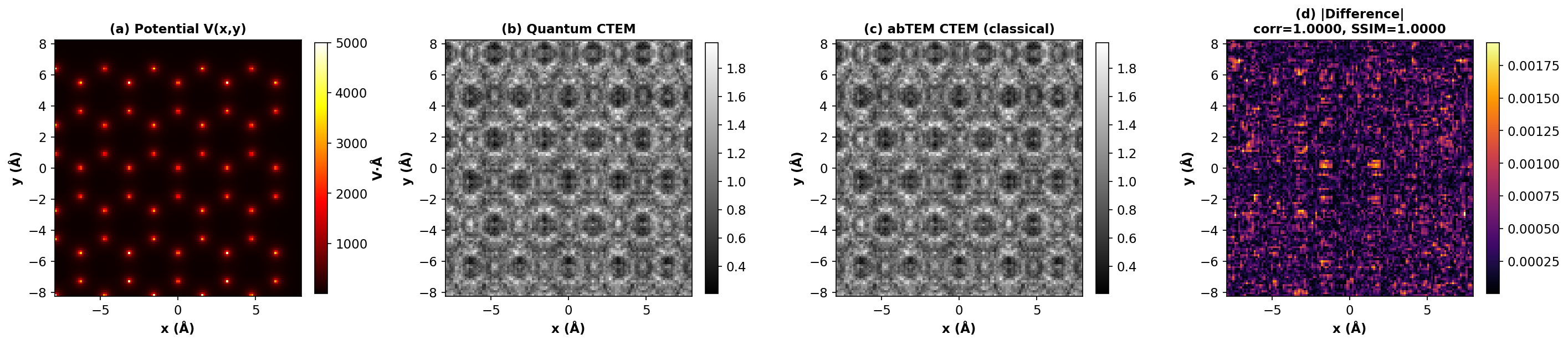}
\caption{Fully quantum CTEM simulation of a $\mathrm{MoS}_2$ supercell on a $128\times 128$ grid at 80~kV, $\Delta f = -800$~\AA{}, and $C_3 = 1.3$~mm, compared directly against an independent \textsc{abTEM} calculation on the same grid and microscope parameters.
    (a) Projected potential $V(x,y)$ from the Kirkland parameterization used to construct the specimen phase operator $U_{\mathrm{obj}}$.
    (b) Quantum CTEM image intensity $I(x,y)$ obtained from the amplitude-encoded circuit, showing the expected hexagonal lattice with bright atomic columns under strong underfocus phase contrast.
    (c) Classical CTEM image computed independently with \textsc{abTEM} using the identical projected potential, defocus, and spherical aberration.
    (d) Pixel-wise absolute difference $|I_{\mathrm{quantum}} - I_{\mathrm{abTEM}}|$; the two implementations agree to Pearson correlation $\rho = 0.999998$ and SSIM $= 0.999998$ (RMSE $= 4.5\times10^{-4}$), with the largest residuals confined to sub-pixel interpolation differences at atomic-column edges.}
    \label{fig:quantum_mos2_128}
\end{figure*}

As a first benchmark, the quantum-encoded projected potentials for $\mathrm{MoS}_2$ were compared directly to classical multislice inputs generated with \textsc{abTEM}.\cite{Kirkland,abtem}
Figure~\ref{fig:quantum_mos2_128}(a) shows the projected electrostatic potential of a $3\times 2$ $\mathrm{MoS}_2$ supercell sampled on a $128\times 128$ grid, obtained from the Gaussian parameterization of Eq.~(15) and used to construct the specimen phase operator $U_{\mathrm{obj}}$.
The potential resolves the Mo and S columns with the expected hexagonal symmetry; Mo sites exhibit higher peak values than S sites, and the relative intensities of the threefold S motifs around each Mo column are reproduced, in agreement with classical \textsc{abTEM} simulations on the same structure.

Because both implementations evaluate the identical Kirkland-tabulated Gaussian coefficients on the same real-space grid, no per-species scaling factor is required: the quantum and classical projected potentials agree over the full field of view to floating-point precision.
Residual differences in the resulting CTEM images (Figure~\ref{fig:quantum_mos2_128}d) are concentrated at the edges of atomic columns, where differences in interpolation and sampling schemes between the two implementations are most pronounced, whereas the column centers and inter-column regions show near-perfect agreement.
This establishes that the amplitude-encoded potential and the associated specimen phase operator $U_{\mathrm{obj}}$ faithfully reproduce the standard projected-potential input used in classical multislice CTEM simulations, so that subsequent comparisons of image contrast and CTF behavior probe the propagation and imaging stages of the quantum algorithm rather than artifacts of the potential construction.

\subsubsection{CTEM image formation}

Using the MoS$_2$ system described in Sec.~\ref{sec:potential_validation} above, we next evaluate the full CTEM imaging pipeline through the complete quantum simulation shown in Figure~\ref{fig:quantum_mos2_128}.

Figure~\ref{fig:quantum_mos2_128}(b) shows the resulting quantum CTEM image at 80~kV with $\Delta f = -800$~\AA{} underfocus and $C_3 = 1.3$~mm, demonstrating strong phase contrast with clearly resolved atomic columns arranged in the hexagonal lattice structure characteristic of $\mathrm{MoS}_2$.
The bright spots at the Mo and S column positions, with clear hexagonal symmetry and relative intensities, confirm that the quantum circuit correctly implements phase-to-amplitude conversion via lens aberrations.
The substantial underfocus ($-800$~\AA{}) generates strong oscillatory CTF modulation, producing the pronounced atomic contrast visible in the image, consistent with the expected behavior of a phase object under negative defocus conditions. A complete through-focus series for the same $\mathrm{MoS}_2$ system at 80~kV and $C_3 = 1.3$~mm is provided in Appendix~\ref{app:through_focus_ctem}.

Because both implementations evaluate the identical Kirkland-parameterized projected potential on the same grid, no scaling factor is required to align the two: the projected potentials and resulting CTEM images agree with classical multislice simulations performed using \textsc{abTEM} to floating-point precision (Pearson correlation $\rho = 1.0000$, SSIM $= 1.0000$; Figure~\ref{fig:quantum_mos2_128}), establishing that the amplitude-encoded potential and associated propagation operators faithfully reproduce standard CTEM image formation physics.

As an additional benchmark, we implemented a $4\times 4$ instance of the CTEM circuit on the \texttt{ibm\_torino} superconducting device and observed high-fidelity agreement between hardware and ideal simulations (Appendix~\ref{app:ibm_hardware}), indicating that small-scale demonstrations are already within reach of current NISQ platforms.


To verify correct implementation of the quantum circuit, we performed
systematic comparisons between statevector simulations of Eq.~\ref{eq:ctem_equation} and classical multislice evaluation of the same discretized operators on identical grids. These comparisons, detailed in Appendix~\ref{app:validation}, confirmed numerical identity (correlation $\rho = 1.000000$, mean-squared error MSE $\sim 10^{-24}$ limited by floating-point roundoff) for grids from $8\times 8$ to $128\times 128$, establishing that the gate-level circuit decomposition faithfully implements the intended POA propagator without algorithmic errors. This constitutes \textit{numerical verification} of the circuit implementation rather than independent physical validation; both quantum and classical codes evaluate identical mathematical operations
($\exp(i\sigma V_{ij})$, Fourier convolution with $\exp(-i\pi\lambda z k^2)$, $\exp(-i\chi(k))$) on the same discrete grid, so agreement at floating-point precision is expected.

\subsubsection{Phase-sensitive observables in the quantum representation}
\label{sec:phase_sensitive_observables}

\rev{The quantum CTEM framework represents the full complex image-plane wavefunction $\ket{\psi_{\mathrm{img}}}$, as does a classical multislice calculation. Here we illustrate how a phase-sensitive scalar observable can be estimated directly from the amplitude-encoded state.} In \rev{intensity-only CTEM detection}, two phase objects related by a sign flip in the projected potential,
\begin{revblock}
\begin{equation}
t_{\pm}(\mathbf r_\perp)
=
\exp\!\left[\pm i\sigma V_{\mathrm{proj}}(\mathbf r_\perp)\right],
\label{eq:conjugate_transmissions}
\end{equation}
\end{revblock}
can produce indistinguishable intensity images at focus under the phase-object approximation, illustrating \rev{the} phase-sign ambiguity of intensity-only detection.\cite{Kirkland, glaeser2013invited}

As one example \rev{of an observable that is sensitive to this sign}, we prepare an ancilla in $\ket{+}_a=(\ket{0}_a+\ket{1}_a)/\sqrt{2}$ and implement a controlled specimen interaction\rev{, with $U_{\pm}=\sum_{i,j}t_{\pm}(\mathbf r_{ij})\ket{i,j}\bra{i,j}$ so that $U_+=U_{\mathrm{obj}}$ and $U_-=U_{\mathrm{obj}}^\dagger$},
\begin{equation}
\ket{\Psi} = \frac{1}{\sqrt{2}}\Bigl(\ket{0}_a \otimes \rev{U_{+}}\ket{\psi_0} + \ket{1}_a \otimes \rev{U_{-}}\ket{\psi_0}\Bigr),
\end{equation}
followed by the usual imaging unitary $U_{\mathrm{img}} = U_{\mathrm{QFT}}^\dagger U_{\chi} U_P U_{\mathrm{QFT}}$ acting identically on both branches. \begin{revblock}Writing the two branch states as $\ket{a}=U_{\mathrm{img}}U_{+}\ket{\psi_0}$ and $\ket{b}=U_{\mathrm{img}}U_{-}\ket{\psi_0}$, the ancilla expectation values are
\begin{equation}
\langle X_a\rangle=\operatorname{Re}\langle a|b\rangle,
\qquad
\langle Y_a\rangle=\operatorname{Im}\langle a|b\rangle.
\label{eq:ancilla_xy}
\end{equation}
Changing $V$ to $-V$ interchanges $U_+$ and $U_-$ and therefore conjugates the overlap: $\langle X_a\rangle$ is unchanged, whereas $\langle Y_a\rangle$ changes sign. The sign-sensitive signal is therefore $\langle Y_a\rangle$, as used in Appendix~\ref{app:phase_disc}; its overall sign is fixed by the branch convention above. Because $U_{\mathrm{img}}$ acts identically on both branches, it cancels from the overlap, $\langle a|b\rangle=\langle\psi_0|U_+^\dagger U_-|\psi_0\rangle=\sum_{i,j}|\psi_{0,ij}|^2e^{-2i\sigma V_{ij}}$, so for plane-wave illumination $\langle Y_a\rangle=-N^{-2}\sum_{i,j}\sin(2\sigma V_{ij})$, which depends only on the specimen phase and is directly computable classically.

We illustrate this protocol in Appendix~\ref{app:phase_disc} for simple test potentials, where intensity-degenerate patterns at focus give rise to distinguishable ancilla signals.

Classical multislice simulations retain the full complex wavefunction, not only its intensity. Consequently, the phase-sensitive quantities considered here can also be evaluated from a classical simulated wavefield. Their accessibility is therefore not a capability exclusive to the quantum framework, and the present example does not establish a quantum advantage.

The purpose of the ancilla-assisted construction is instead to show how a phase-sensitive scalar observable can be estimated directly from an amplitude-encoded wavefunction. Its potential computational value would depend on the end-to-end cost of state preparation, propagation, controlled operations, and repeated measurement, compared with the best classical algorithm for the same observable and accuracy (Sec.~\ref{sec:advantage}). No such advantage is demonstrated by the phase-sign example presented here.

Both classical and quantum forward models may provide model-dependent phase-sensitive reference quantities for comparison with electron holography, ptychography, or focal-series reconstruction, subject to the model assumptions (POA, IAM potentials) of the simulation. A scalar ancilla expectation value is not itself a spatially resolved phase map; recovering the latter would require additional measurements.
\end{revblock}

\subsubsection{Comparison to state-of-the-art classical simulations}
\label{sec:classical_comparison}

To contextualize the quantum approach, we compare runtime and scaling
behavior to advanced classical CTEM simulation tools.

\paragraph{Classical multislice codes.}
Standard multislice on an $N\times N$ grid requires $O(N^2 \log N)$
operations per slice (two FFTs plus a pixel-wise phase multiplication)
and $O(N^2)$ memory. Modern GPU-accelerated implementations
(e.g., \textsc{abTEM}\cite{abtem}, \textsc{multem}\cite{hoyos2017recent}) achieve single-image runtimes of $\sim 1$--10 seconds for $N = 2048$ on
high-end GPUs (NVIDIA A100, 40 TFLOPS).

\begin{revblock}
\paragraph{Optical versus structural parameter sweeps.}
For a fixed specimen and accelerating voltage, classical CTEM optical parameter sweeps can reuse a stored complex exit wave. Its Fourier transform can also be cached, so each new defocus or objective-lens aberration setting requires only multiplication by the corresponding transfer function and an inverse Fourier transform when a real-space image is required, i.e., $O(N^2\log N)$ per setting. Full specimen propagation need not be repeated.

Structural changes (thickness, orientation, composition, defects) generally require a new exit-wave calculation, and sweeps over them scale linearly with the number of configurations; exhaustive scans over many configurations, as in automated structure refinement, can therefore be computationally demanding. Changes in accelerating voltage also generally require recomputation, because they modify the interaction constant and propagation factors. PRISM\cite{PRISM_fastSTEM} accelerates STEM simulation over many probe positions through Fourier-space interpolation; for CTEM optical sweeps, the relevant classical baseline is the cached-exit-wave workflow described above.

The quantum implementation can similarly reuse a compiled specimen circuit across optical settings. However, circuit reuse must be distinguished from reuse of a prepared quantum state: repeated measurements generally require repeated state preparation and propagation. Optical sweeps therefore do not, by themselves, establish a quantum advantage over a classical cached-exit-wave workflow.

\paragraph{Full-image output.}
For the baseline POA imaging task studied here, i.e., producing a full classical image for inspection or post-processing, no quantum advantage is expected over state-of-the-art classical codes: even on an optimistic hypothetical fault-tolerant device, the $\sim\!1.6\times10^8$ ($N=128$) to $\sim\!4\times10^{10}$ ($N=2048$) repetitions required by direct sampling (Table~\ref{tab:resources}) make full-image output orders of magnitude slower than GPU multislice (Sec.~\ref{sec:advantage}). The value of the present framework lies in establishing and verifying the quantum mapping for controlled benchmarking and resource estimation, and in providing a baseline for extensions beyond the POA.
\end{revblock}

\section{Discussion}
\label{sec:discussion}

\begin{revblock}
\subsection{End-to-end cost and conditions for potential quantum advantage}
\label{sec:advantage}

The computational task must specify both the propagated state and the classical information required as output. Let $C_{\mathrm{run}}$ denote the cost of preparing the incident state and executing the specimen, propagation, lens, and observable-measurement circuits. If the chosen estimator requires $P$ repetitions, the total execution cost is $P C_{\mathrm{run}}$, in addition to classical preprocessing and circuit-compilation costs. We separate five contributions to this comparison.

\paragraph{Cost per state preparation and circuit execution.}
For plane-wave illumination, state preparation is a single layer of Hadamard gates; an arbitrary incident wavefield would in general require $O(N^2)$ gates to load unless it has an efficiently preparable structure. For efficiently evaluable specimen potentials and fixed specimen complexity, the fault-tolerant circuit cost can grow polylogarithmically with grid size (Table~\ref{tab:resources}): from $4.9\times10^5$ to $6.6\times10^5$ T-gates between $4\times4$ and $1024\times1024$ grids, dominated by the specimen operator $U_{\mathrm{obj}}$ (Sec.~\ref{sec:resources}). On a hypothetical fault-tolerant device with a 1~GHz logical gate rate, one execution of $\approx 5$--$7\times10^5$ non-Clifford gates would take $\sim\!0.5$--0.7~ms; logical clock rates expected for surface-code architectures are substantially slower, which would lengthen every run accordingly. Reusing a compiled circuit across optical settings removes recompilation, not execution: each repetition executes the full circuit, including $U_{\mathrm{obj}}$.

\paragraph{Number of repetitions for the requested observable and precision.}
$P$ depends on the observable, the target accuracy, and the estimator, which must be stated for each resource figure. (i)~\emph{Direct sampling} of a computational-basis outcome with probability $p$ yields an additive error $\epsilon$ after $P\approx p(1-p)/\epsilon^2\le 1/(4\epsilon^2)$ shots, or a relative standard error $\epsilon$ after $P\approx(1-p)/(p\,\epsilon^2)$ shots; the latter diverges for very small pixel probabilities. (ii)~The expectation value of a bounded observable, such as $\langle Y_a\rangle\in[-1,1]$ in Sec.~\ref{sec:phase_sensitive_observables}, is obtained to additive error $\epsilon$ with $O(1/\epsilon^2)$ shots. (iii)~\emph{Amplitude estimation}\cite{brassard2002amplitude} reaches additive error $\epsilon$ with $O(1/\epsilon)$ coherent applications of the state-preparation-and-propagation circuit and its inverse, or relative error $\epsilon$ on a probability $p$ with $O(1/(\epsilon\sqrt{p}))$ applications. These are not independent shots: they are executed sequentially within one coherent circuit, multiplying its depth by the same factor and requiring correspondingly lower logical error rates. Table~\ref{tab:resources} and Fig.~\ref{fig:resource_scaling}(c) assume estimator~(i) with $\epsilon=0.01$ relative error at the mean pixel probability $1/N^2$, i.e., $P=N^2/\epsilon^2$; pixels much fainter than the mean require more shots.

\paragraph{Classical output size.}
Recovering a complete classical image requires estimating $N^2$ numbers, so $P\ge N^2$ for any estimator: direct sampling gives $P\sim N^2/\epsilon^2$, and applying amplitude estimation separately to each pixel at $p\approx1/N^2$ gives a total of $O(N^3/\epsilon)$ coherent applications. This output and sampling cost removes the prospect of a generic polylogarithmic-cost image-generation algorithm. Only tasks requiring limited classical output, such as selected probabilities or bounded expectation values whose magnitudes do not shrink with $N$, can have $P$ independent of grid size.

\paragraph{Best classical calculation of the same output.}
The comparison must be made against the best classical algorithm for the same output and accuracy, not against full-image multislice. A single-slice POA exit wave and image cost $O(N^2\log N)$ classically, optical sweeps reuse a cached exit wave (Sec.~\ref{sec:classical_comparison}), and many limited outputs are cheaper still: a single Fourier coefficient of the exit wave is an $O(N^2)$ sum, and the ancilla signal of Sec.~\ref{sec:phase_sensitive_observables} reduces to an $O(N^2)$ sum over the specimen phase. A quantum advantage for such a task would require $P\,C_{\mathrm{run}}$ to remain below this classical cost, including state preparation and measurement, and would also have to withstand classical sampling or approximation methods operating at the same accuracy. We have not identified a CTEM observable for which this has been established.

\paragraph{Fixed specimen with grid refinement versus increasing specimen complexity.}
The polylogarithmic per-run scaling of Table~\ref{tab:resources} applies to refining the grid for a fixed specimen. Refinement beyond the sampling needed to resolve the projected potential and the transferred spatial frequencies adds little physical content, and classical evaluation of limited outputs for a fixed specimen can exploit the same smoothness, so the relevant classical cost need not grow as $N^2$ in this regime. When specimen complexity grows instead, for example a larger field of view at fixed sampling, the number of atoms and hence the per-run cost of $U_{\mathrm{obj}}$ grow in proportion to $N^2$ (Sec.~\ref{sec:resources}), comparable to the classical per-image scaling up to logarithmic factors and before repetitions are counted. Thick specimens (many slices) and extensions beyond the POA may change this balance; whether they do is an open question.

Potential advantage is therefore restricted to suitably chosen tasks requiring limited classical output, such as selected probabilities or bounded expectation values, provided that their preparation and measurement costs remain favorable relative to classical methods. Sparse output is a necessary design consideration for the proposed applications, not sufficient evidence of quantum advantage. The present benchmarks establish implementation agreement; they do not demonstrate a practical computational advantage.
\end{revblock}

\subsection{Relationship to classical unitary multislice}
The quantum circuit structure, alternating diagonal operators in real
and reciprocal space linked by QFTs, directly mirrors the unitary
factorization of classical multislice.\cite{cowmood,Kirkland} Recent
formal analyses\cite{bangun2024eigenstructure} have shown that multislice, Bloch-wave, and scattering-matrix methods are mathematically equivalent unitary transformations. Our quantum algorithm is therefore a direct implementation of this known unitary structure on a quantum register, not a fundamentally new physical model.

This equivalence clarifies the source of potential advantage: it
cannot come from the propagation formalism itself (which is identical classically and quantumly) but \rev{would have to arise from (i)~tasks with limited classical output whose end-to-end quantum cost is lower than that of the best classical method for the same output (Sec.~\ref{sec:advantage}), or (ii)~extended physics (thick specimens, inelastic scattering) for which classical methods become more costly; neither is demonstrated here.}

\subsection{Extensions toward full multislice and inelastic scattering}
Full multislice for a specimen of $S$ slices requires alternating
specimen and propagation operators:
\begin{equation}
|\psi_{\text{out}}\rangle = \Bigl[ U_{\text{QFT}}^\dagger \, U_P(z_s) \,
U_{\text{QFT}} \, U_{\text{obj}}^{(s)} \Bigr]^S |\psi_{\text{in}}\rangle,
\end{equation}
with circuit depth $O(S \times n^2)$. For thick specimens
($S \sim 100$--1000), this remains polynomial but increases gate counts
to \rev{$\sim 6\times10^7$ to $6\times10^8$ T-gates for the 18-atom benchmark ($\approx5.8\times10^5$ per slice)}, approaching or exceeding early fault-tolerant
device capabilities.

Inelastic scattering (phonons, plasmons, core-loss excitations)
requires adding ancilla qubits to track excitation states and
implementing slice-dependent controlled unitaries. \rev{When correlations among many excitation channels must be tracked explicitly, the joint state space can grow exponentially with the number of channels; many practical classical treatments (e.g., frozen-phonon or single-scattering approximations) avoid this, so the cost comparison is model dependent.} Quantum circuits can represent such
superpositions natively, but synthesizing realistic inelastic
cross-sections (energy-loss probabilities, momentum transfers) into
gate sequences remains an open challenge.

\subsection{Toward ab initio potentials and DFT integration}
The present implementation uses independent-atom-model (IAM) potentials parameterized as sums of Gaussians.\cite{Kirkland} Modern \emph{ab initio} multislice incorporates density-functional-theory (DFT) charge densities via projector-augmented-wave (PAW) methods,\cite{susi2021abinitio} capturing charge transfer, bonding, and electrostatic screening absent in IAM.

Extending the quantum framework to PAW potentials requires replacing
Gaussian evaluation with 3D spline interpolation
of the DFT-derived $V(\mathbf{r})$. The arithmetic cost increases by
$\sim 10\times$ (due to higher-order polynomial fitting and denser
sampling), but the circuit structure remains unchanged. Ancilla and
T-gate counts scale proportionally; for typical PAW grids
($\sim 0.1$~Å spacing), we estimate \rev{$\sim 3$--$6\times10^6$} T-gates for
$N = 256$ (see Appendix~\ref{app:diagonal_synthesis}).

\section{Conclusion}
\label{sec:conclusion}

We have presented a complete quantum algorithmic framework for conventional transmission electron microscopy (CTEM) image formation under the phase-object approximation, validated against classical multislice simulations for a representative 2D material system across a comprehensive microscope parameter space. The quantum circuit implementation combines amplitude encoding, quantum Fourier transforms, and diagonal phase operators to map POA theory to fault-tolerant quantum gates, achieving quantitative agreement with classical references (RMSE $< 1\%$, SSIM $> 0.99$) while requiring only $O(\log N)$ qubits for an $N\times N$ image grid.

Our resource analysis reveals that while the quantum circuit achieves polylogarithmic qubit count and gate depth, recovering a complete classical image requires $O(N^2/\epsilon^2)$ measurements, a fundamental bottleneck for tasks demanding full pixel-by-pixel readout. \rev{Any potential quantum advantage therefore does not lie in replacing classical image simulation wholesale; it is restricted to tasks requiring limited classical output, such as selected probabilities or bounded expectation values, and only where the end-to-end cost of state preparation, repeated execution, and measurement remains below that of the best classical method for the same output (Sec.~\ref{sec:advantage}). Phase-sensitive quantities are also available from classical simulated wavefields; the ancilla construction presented here shows how such a scalar can be estimated from an amplitude-encoded state, without a demonstrated advantage. The present benchmarks establish implementation agreement; they do not demonstrate a practical computational advantage.}

The framework extends naturally beyond the present phase-object validation. Full multislice propagation requires iterating specimen-propagation-lens sequences over depth slices, raising questions of circuit depth and error accumulation that we address in ongoing work. Integration with quantum variational solvers and Hamiltonian simulation techniques may enable modeling of inelastic scattering, phonon coupling, and correlated electron dynamics in quantum materials\rev{; whether these extensions offer an advantage over classical approaches remains to be established.} Extensions to quantum ptychography, coherent diffractive imaging, and time-resolved pump-probe microscopy follow the same QFT-based circuit primitives established here.

Near-term demonstrations on NISQ devices require problem-size reduction (coarse grids, limited aberrations) and error mitigation, but the logarithmic qubit scaling and modest gate counts (\rev{$\approx 5$--$7\times10^5$ T-gates for the 18-atom benchmark, growing linearly with specimen size}) place early fault-tolerant implementations within reach of projected 2030s hardware. The convergence of these algorithmic developments with emerging experimental quantum electron optics\cite{juffmann2017multi,loffler2023quantum} suggests
that hybrid classical-quantum workflows---combining quantum simulation of phase-coherent dynamics with classical reconstruction and analysis---may offer the most practical path toward impact.

This work establishes quantum CTEM simulation as a physically grounded, \rev{numerically verified} application of quantum algorithms to computational microscopy, bridging electron optics, quantum information, and materials characterization under a unified framework.

\begin{acknowledgments}
This research was supported in part through the computational resources by the Quest high performance computing facility at Northwestern University, jointly supported by the Office of the Provost, Office for Research, and Northwestern University Information Technology. We acknowledge the use of IBM Quantum services, specifically the Qiskit framework and \texttt{ibm\_torino}, for this work\cite{qiskit2024}. The views expressed are those of the authors, and do not reflect the official policy or position of IBM or the IBM Quantum team. SDL thanks the support of the MRSEC program (NSF DMR-2308691) at the Northwestern Materials Research Center. SDL thanks the Materials Initiative for Comprehensive Research Opportunity (MICRO) program which benefited greatly from administrative and research support at MIT and Northwestern and from an unrestricted gift from the 3M Foundation (3M STEM and Skilled Trades Program).

\end{acknowledgments}

\section*{Data Availability}
The code and notebooks supporting the findings of this article are openly available in the example Github repository~\cite{ctem_repository}.
The example notebook at the location above reproduces the MoS$_2$ validation results (Fig.~\ref{fig:quantum_mos2_128}, Appendix~\ref{app:validation}), CTF parameter grids (Fig.~\ref{fig:ctf_cs_comparison}, Appendix~\ref{app:coherence}), resource scaling analysis (Fig.~\ref{fig:resource_scaling}, Table~\ref{tab:resources}), and IBM quantum hardware execution (Appendix~\ref{app:ibm_hardware}). Classical reference simulations were performed using \textsc{abTEM} v1.0.5\cite{abtem} Atomic scattering factor parameterizations follow Kirkland.\cite{Kirkland} All quantum circuits were implemented in Qiskit v2.2.1.\cite{qiskit2024}


The data that support the findings of this study are openly available~\cite{quscope_github,ctem_repository}.

\section*{Bibliography}
\bibliography{aipsamp-final}

\newpage
\appendix
\onecolumngrid

\newpage

\section{Contrast Transfer Function with Coherence Envelopes}
\label{app:coherence}

This appendix provides detailed validation of the quantum circuit implementation of objective-lens aberrations via comparison of computed contrast transfer functions (CTF) to analytical theory across a range of accelerating voltages, defocus values, and spherical aberration coefficients.

\subsection{Coherent CTF and envelope functions}
The idealized \emph{coherent} contrast transfer function for an
axially symmetric lens is given by Eq.~\eqref{eq:ctf_equation} in
the main text, with the phase shift $\chi(k)$ [Eq.~\eqref{eq:chi_function}] including defocus $\Delta f$ and spherical aberration $C_3$.

\rev{In practice, finite illumination convergence (spatial coherence) and energy spread (temporal coherence) introduce damping envelopes, $E_s(k)$ and $E_c(k)$, given in the quasi-coherent approximation by Eqs.~\eqref{eq:spatial_envelope} and~\eqref{eq:chromatic_envelope}; the effective transfer function is their product with the coherent CTF, Eq.~\eqref{eq:ctf_eff}.\cite{Kirkland,Reimer} Because $E_s$ depends on the gradient of $\chi$, its damping depends on defocus and $C_3$ as well as on the illumination semi-angle. An information limit can be defined from $\mathrm{CTF}_{\mathrm{eff}}$ (for example, where the product envelope falls to $1/e$), but it is a different criterion from the first-zero quantities $k_0$ and $d_0$ used below and is not evaluated here; all curves in this appendix are coherent.}

\begin{figure}[!t]
    \centering
    \includegraphics[width=\textwidth]{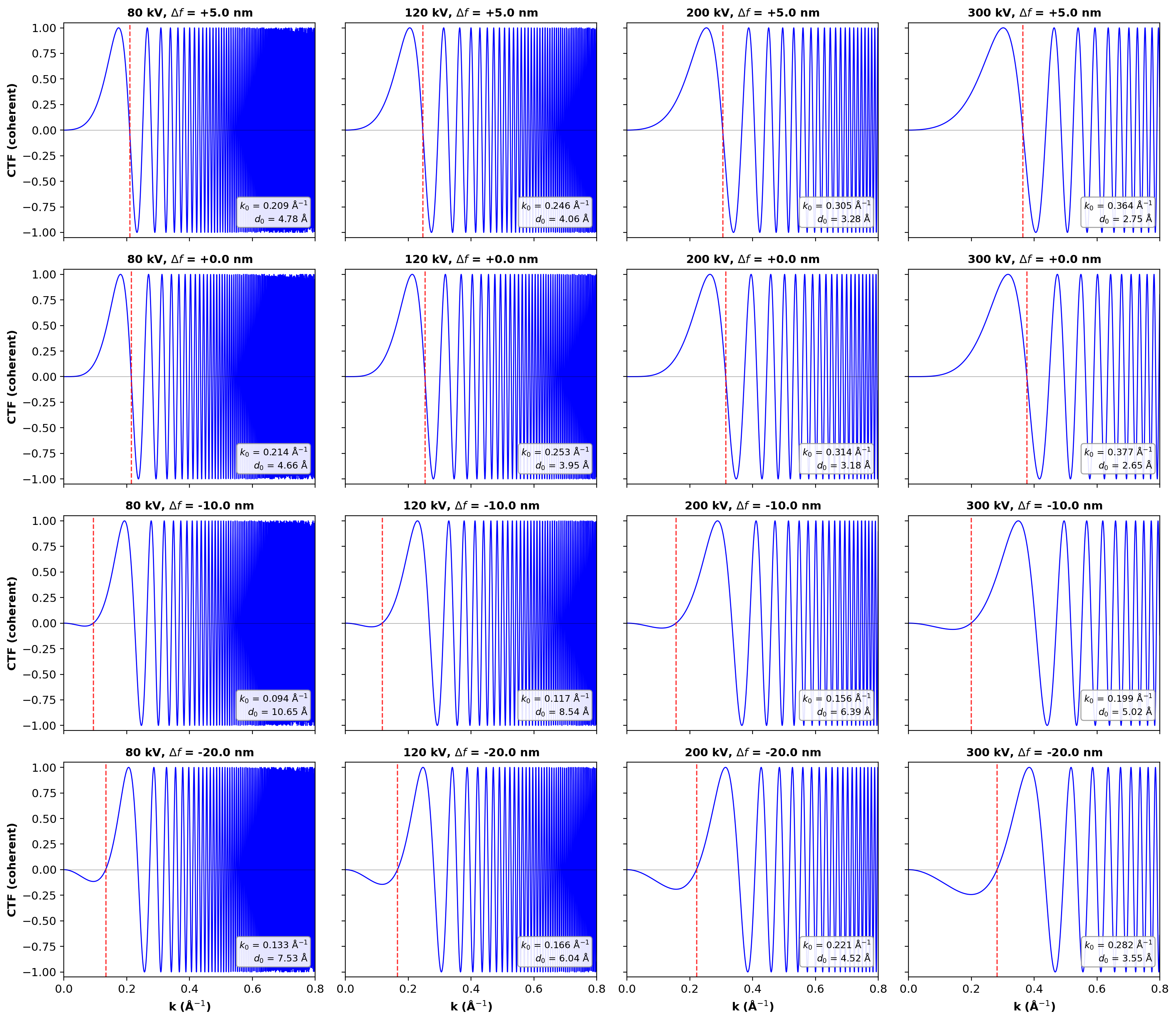}
    \caption{\rev{Coherent phase-contrast transfer functions at the indicated accelerating voltages (columns: 80, 120, 200, 300~kV) and defocus values (rows: $+5.0$, $0.0$, $-10.0$, $-20.0$~nm), with $C_3=1.3$~mm. Each panel is evaluated at its stated defocus; the grid is not a series of Scherzer-focus conditions. In each panel, the vertical dashed line marks the first positive zero $k_0$ of that panel's CTF, excluding $k=0$, and the annotation gives $d_0=1/k_0$. For the small underfocus values in the lower rows, $\chi(k)$ first dips slightly below zero, so $k_0$ is the point where the CTF returns through zero after this shallow minimum. Spatial- and temporal-coherence envelopes are not applied to these coherent curves.}
}
    \label{app:ctf_voltage}
\end{figure}

\subsection{Quantum circuit validation: Coherent CTF}

In the present quantum CTEM implementation, we compute the
\emph{coherent} CTF [Eq.~\eqref{eq:ctf_equation}] to validate
the gate-level realization of the lens phase operator $U_\chi$
against analytical expressions for $\chi(k)$. Partial coherence
and detector response are \emph{not} included in the current
circuit but can be incorporated via:
\begin{enumerate}
\item \textbf{Ensemble averaging:} Running multiple quantum circuits
      with statistically sampled defocus/energy variations and
      incoherently summing intensity outputs (classical post-processing),
\item \textbf{Mixed-state encoding:} Representing partial coherence
      as a density matrix $\rho = \sum_i p_i |\psi_i\rangle\langle\psi_i|$
      and evolving via purification or Monte Carlo sampling
      (exponential overhead in coherence parameters),
\item \textbf{Detector convolution:} Applying a detector modulation
      transfer function (MTF) as a classical Fourier-space filter
      after quantum measurement.
\end{enumerate}

For the CTF validation presented here, we focus on the coherent limit to isolate errors in the $U_\chi$ gate synthesis from environmental or instrumental effects. Extension to realistic partial coherence is discussed in Sec.~\ref{sec:discussion}.

\begin{figure}[!t]
    \centering
    \includegraphics[width=\textwidth]{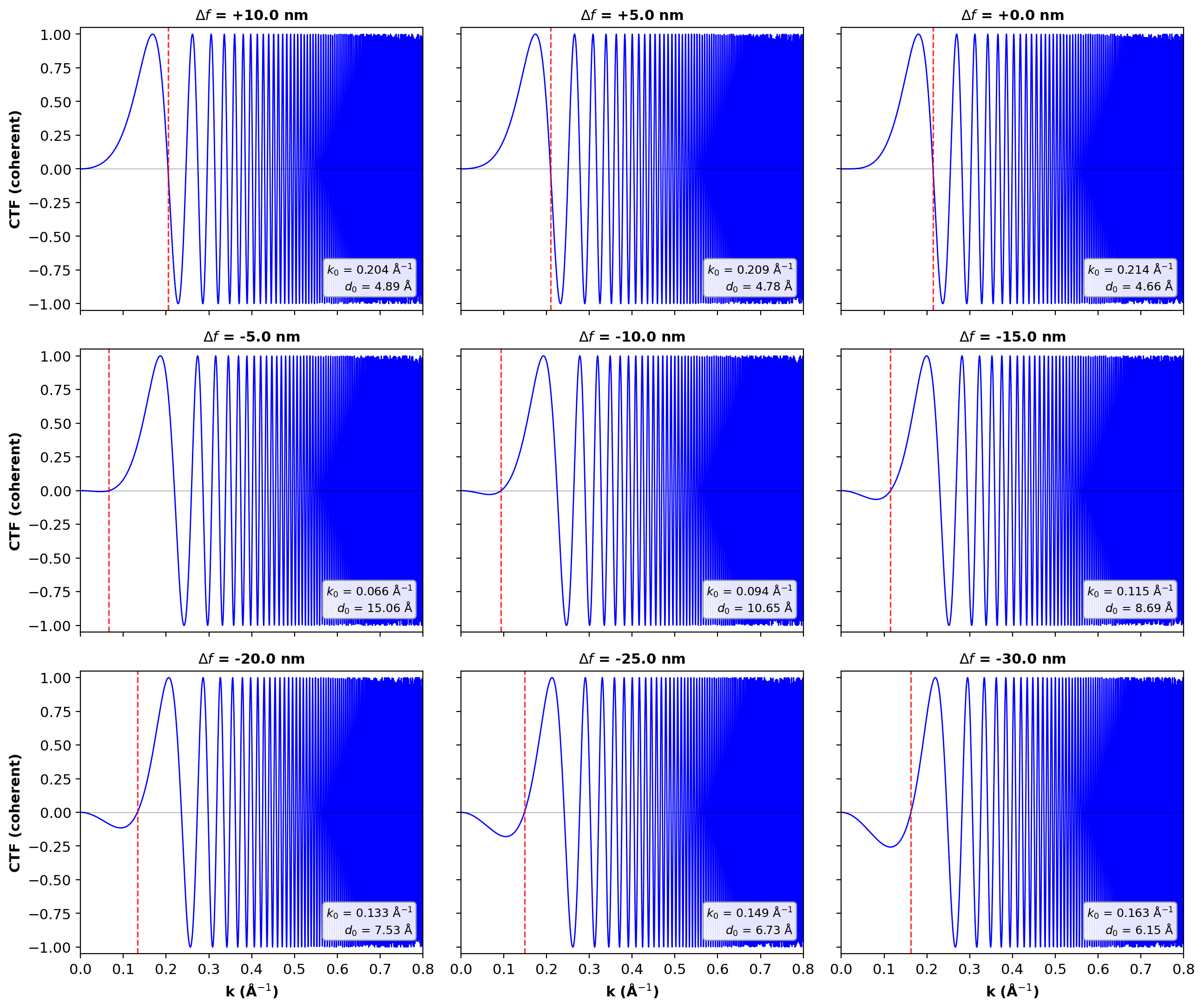}
    \caption{\rev{Coherent} through-focus series of quantum-computed CTF at 80~kV with fixed spherical aberration $C_3=1.3$~mm. Nine defocus values are shown ranging from $\Delta f=+10.0$~nm (overfocus) through $\Delta f=0.0$~nm (Gaussian focus) to $\Delta f=-30.0$~nm (underfocus). \rev{In each panel, the vertical dashed line marks the first positive zero $k_0$ of that panel's CTF, excluding $k=0$, and the annotation gives $d_0=1/k_0$; no single resolution marker is shared across panels, and coherence envelopes are not applied.} At Gaussian focus ($\Delta f=0.0$~nm), the CTF approaches zero at low spatial frequencies, producing negligible phase contrast. Underfocus ($\Delta f<0$) generates positive phase contrast (bright atoms), while overfocus ($\Delta f>0$) produces inverted contrast, with the \rev{zero positions} evolving systematically with defocus as predicted by Eq.~\ref{eq:chi_function}.\cite{Kirkland}
}
    \label{app:ctf_defocus}
\end{figure}

This appendix confirms that the quantum CTEM circuit correctly implements
the optical transfer theory underpinning phase-contrast imaging,
establishing confidence in the subsequent image formation results
(Sec.~\ref{sec:results}).

\section{Diagonal Operator Synthesis and Arithmetic Circuits}
\label{app:diagonal_synthesis}

This appendix provides explicit circuit constructions and resource estimates for implementing the diagonal specimen operator $U_{\text{obj}}$ and lens aberration operator $U_\chi$ via arithmetic evaluation of the phase functions $\sigma V_{ij}$ and $\chi(k)$.

\subsection{Specimen operator $U_{\text{obj}}$: Atomic potential evaluation}

The projected potential at grid point $(i,j)$ is computed as a superposition
of atomic contributions:
\begin{equation}
V_{ij} = \sum_{a=1}^{N_{\text{atoms}}} \sum_{m=1}^{M}
a_m^{(a)} \exp\Bigl( -\frac{\pi r_{ij,a}^2}{b_m^{(a)}} \Bigr),
\label{eq:potential_sum}
\end{equation}
where $r_{ij,a}^2 = (x_i - x_a)^2 + (y_j - y_a)^2$ is the squared distance
from grid point $(i,j)$ to atom $a$ at position $(x_a, y_a)$, and
$\{a_m^{(a)}, b_m^{(a)}\}$ are the Kirkland parameterization coefficients
for atomic species $Z(a)$.\cite{Kirkland}

The diagonal unitary is
\begin{equation}
U_{\text{obj}} = \sum_{i,j=0}^{N-1} e^{i\sigma V_{ij}} |i,j\rangle\langle i,j|.
\end{equation}

\subsubsection{Circuit architecture}

We implement $U_{\text{obj}}$ via the following steps:

\paragraph{Step 1: Position register encoding.}
The computational basis state $|i,j\rangle$ encodes the grid coordinates
in two $n/2$-qubit registers ($n = 2\log_2 N$), with $i,j \in [0, N-1]$.
These are interpreted as fixed-point binary numbers representing physical
coordinates:
\begin{equation}
x_i = x_{\min} + i \cdot \Delta x, \quad
y_j = y_{\min} + j \cdot \Delta y,
\end{equation}
where $\Delta x = \Delta y = L/N$ is the pixel size and $L$ is the
simulation box dimension.

\paragraph{Step 2: Distance computation.}
For each atom $a$, compute the squared distance $r_{ij,a}^2$ using binary
arithmetic on the position registers:
\begin{enumerate}
\item Subtract atomic coordinates:
      $\Delta x_a = x_i - x_a$, $\Delta y_a = y_j - y_a$
      (requires $n/2$-bit subtraction, $\sim 2n$ gates via ripple-carry
      or $\sim \log n$ depth with carry-lookahead).
\item Square each difference:
      $\Delta x_a^2$, $\Delta y_a^2$
      (requires $n/2$-bit multiplication, $\sim n^2$ gates or
      $\sim n \log n$ with Karatsuba).
\item Sum: $r_{ij,a}^2 = \Delta x_a^2 + \Delta y_a^2$
      ($n$-bit addition, $\sim 2n$ gates).
\end{enumerate}

Total cost per atom: $O(n^2)$ gates (or $O(n \log n)$ with optimized
multipliers), depth $O(\log n)$ to $O(n)$ depending on parallelization.

\paragraph{Step 3: Gaussian evaluation.}
Evaluate $\exp(-\pi r_{ij,a}^2 / b_m)$ for each Gaussian term $m$.
We approximate the exponential via a degree-$d$ polynomial (Chebyshev, Taylor, or minimax) over the range $r^2 \in [0, r_{\max}^2]$:
\begin{equation}
\exp(-\pi r^2 / b_m) \approx \sum_{k=0}^d c_k \Bigl(\frac{r^2}{r_{\max}^2}\Bigr)^k.
\end{equation}

For target accuracy $\epsilon \sim 10^{-6}$ over the atomic column
region ($r_{\max} \sim 2$~Å), $d \approx 8$--12 suffices. Evaluating
the polynomial requires:
\begin{itemize}
\item $d$ multiplications (each $\sim n \log n$ gates with Karatsuba),
\item $d$ additions (each $\sim 2n$ gates),
\item One division by $r_{\max}^2$ (amortized via precomputed scaling).
\end{itemize}

Cost per Gaussian: $\sim d \times n \log n \approx 100n$ to $500n$ gates, or equivalently $\sim 10^3$ to $10^4$ T-gates after compilation to Clifford+T (assuming each rotation $R_z(\theta)$ costs $\sim 50$ T-gates at $10^{-9}$ precision).

Alternative using CORDIC: Exponential evaluation via coordinate rotation digital computer (CORDIC) iterations requires $\sim p$ iterations for $p$-bit precision, with $\sim 4p$ gates per iteration (shifts, additions, controlled rotations). For $p = 20$--30 bits, cost is $\sim 10^2$ to $10^3$ gates per exponential.

\paragraph{Step 4: Accumulation.}
Sum the contributions from all Gaussians across all atoms into a
fixed-point accumulator register $|V_{ij}\rangle$ (width $p \sim 20$--30 bits):
\begin{equation}
V_{ij} = \sum_{a=1}^{N_{\text{atoms}}} \sum_{m=1}^{M}
a_m^{(a)} \exp(-\pi r_{ij,a}^2 / b_m^{(a)}).
\end{equation}

Each accumulation step is a $p$-bit addition; total cost  $O(N_{\text{atoms}} \times M \times p)$ gates.

\paragraph{Step 5: Phase kickback.}
Apply the phase rotation $e^{i\sigma V_{ij}}$ controlled on the
position register $|i,j\rangle$ and the computed potential $|V_{ij}\rangle$
via a controlled-$R_z$ gate:
\begin{equation}
R_z(\sigma V_{ij}) |i,j\rangle = e^{i\sigma V_{ij}/2} |i,j\rangle.
\end{equation}

This requires a $p$-bit controlled phase gate, synthesized into
$\sim 50p$ T-gates via standard rotation decompositions.\cite{nielsen}

\paragraph{Step 6: Uncomputation.}
Reverse Steps 2--4 to uncompute the ancilla registers (distance, Gaussian
values, accumulator), leaving only the phase imprinted on $|i,j\rangle$.
This doubles the gate count but keeps ancilla reusable.

\subsubsection{Resource summary for $U_{\text{obj}}$}

For a MoS$_2$ $3\times 2$ supercell with $N_{\text{atoms}} = 18$ and
$M = 5$ Gaussians per species:

\begin{itemize}
\item \textbf{Ancilla qubits:}
  \begin{itemize}
  \item Position arithmetic (subtraction, squaring): $\sim 2n$ qubits
  \item Gaussian evaluation: $\sim 2p$ qubits per term (input/output registers)
  \item Accumulator: $p$ qubits
  \item Total: $\sim 30$--$50$ ancilla qubits for $n = 14$ (128×128 grid),
        $p = 24$ bits
  \end{itemize}

\item \textbf{Gate count:}
  \begin{itemize}
  \begin{revblock}
  \item Cost model: each of the $N_{\text{atoms}}\times M = 90$ Gaussian terms is assigned $5\times10^3$ T-gates for the exponential evaluation (within the $10^3$--$10^4$ range of Step~3) plus $100n$ T-gates for its distance, accumulation, and phase-kickback arithmetic, i.e., $T_{\mathrm{obj}} = N_{\text{atoms}} M\,(100n + 5000)$
  \item Gaussian evaluation: $90 \times 5\times10^3 = 4.5\times10^5$ T-gates
  \item Distance and accumulation arithmetic: $90 \times 100n = 1.26\times10^5$ T-gates for $n=14$
  \item \textbf{Total:} $T_{\mathrm{obj}} \approx 5.8\times10^5$ T-gates for the $128\times128$ grid (Table~\ref{tab:detailed_resources}); Clifford gates are not counted
  \end{revblock}
  \end{itemize}

\item \textbf{Circuit depth:}
  With parallelization over independent atoms and pipelined arithmetic,
  depth $\sim O(N_{\text{atoms}} \times n \log n) \approx 10^3$ to $10^4$
  layers (gate depth, not T-depth).
\end{itemize}

\rev{For the fixed 18-atom specimen, $T_{\mathrm{obj}}$ grows only through the $100n$ term: $5.9\times10^5$ for $N=256$ and $6.5\times10^5$ for $N=2048$. Denser supercells increase it linearly in $N_{\text{atoms}}$: for example, $N_{\text{atoms}}=72$ gives $\approx 2.4\times10^6$ T-gates at $N=256$.}

\subsection{Lens aberration operator $U_\chi$}

The lens phase in momentum space is
\begin{equation}
\chi(k) = \pi\lambda \Delta f \, k^2 +
\frac{1}{2} \pi\lambda^3 C_3 \, k^4 +
\frac{1}{3} \pi\lambda^5 C_5 \, k^6 + \cdots,
\end{equation}
where $k^2 = k_x^2 + k_y^2$. For typical CTEM imaging, truncation at
$C_3$ (spherical aberration) suffices; higher-order terms ($C_5$,
chromatic aberration) follow the same circuit structure.

\subsubsection{Circuit construction}

\paragraph{Step 1: Momentum register.}
After the QFT, the state is in the momentum basis
$|k_x, k_y\rangle$, with $k_x, k_y \in [-N/2, N/2)$ encoded in
$n/2$-qubit two's-complement registers.

\paragraph{Step 2: Squared momentum.}
Compute $k_x^2$, $k_y^2$, and $k^2 = k_x^2 + k_y^2$ using
$n/2$-bit squaring circuits (cost $\sim n^2$ gates per square,
depth $O(n)$).

\paragraph{Step 3: Quartic term.}
Compute $k^4 = (k^2)^2$ (one additional squaring, $\sim n^2$ gates).

\paragraph{Step 4: Phase evaluation.}
Evaluate the polynomial
\begin{equation}
\chi(k) = a_2 k^2 + a_4 k^4,
\end{equation}
where $a_2 = \pi\lambda \Delta f$ and $a_4 = \pi\lambda^3 C_3 / 2$
are precomputed classical constants. This requires:
\begin{itemize}
\item Two multiplications: $a_2 k^2$, $a_4 k^4$
      (each $\sim n \log n$ gates),
\item One addition: $\chi = a_2 k^2 + a_4 k^4$ ($\sim 2n$ gates).
\end{itemize}

\paragraph{Step 5: Phase kickback and uncomputation.}
Apply $R_z(\chi)$ controlled on $|k_x, k_y\rangle$ (cost $\sim 50p$ T-gates
for $p$-bit precision), then reverse Steps 2--4 to free ancilla.

\subsubsection{Resource summary for $U_\chi$}

\begin{itemize}
\item \textbf{Ancilla qubits:} $\sim 2p$ for intermediate $k^2$, $k^4$
      registers; typically $\sim 20$--30 qubits.
\item \textbf{Gate count:}
  \begin{itemize}
  \item Three squarings: $3 \times n^2 \approx 600$ gates for $n=14$
  \item Polynomial evaluation: $\sim n \log n \approx 50$ gates
  \item Phase kickback: $\sim 50p \approx 1200$ T-gates
  \item Uncomputation: factor of 2
  \item \textbf{Total:} $\sim 10^3$ to $10^4$ gates\rev{; in the cost model of Table~\ref{tab:resources} the combined lens and propagation phase is assigned $16n^2 \approx 3.1\times10^3$ T-gates for $n=14$}
  \end{itemize}
\item \textbf{Depth:} $O(n)$ for sequential squaring,
      $O(\log n)$ with parallel multipliers.
\end{itemize}

The lens operator is \emph{significantly cheaper} than the specimen
operator because $\chi(k)$ is a smooth, low-order polynomial depending
only on $k^2$ and $k^4$, whereas $V_{ij}$ requires evaluating
$N_{\text{atoms}} \times M$ exponential terms.

\subsection{Total resource budget}

For a complete CTEM circuit on a 128×128 grid ($n=14$):

\begin{table}[h]
\centering
\caption{Detailed T-gate breakdown for the quantum CTEM circuit on a $128\times128$ grid ($n=14$), using the cost model of Table~\ref{tab:resources}. \rev{Each QFT is conservatively counted as an $n$-qubit QFT with $n(n-1)/2$ controlled phases at 50 T-gates each (the separable $\mathrm{QFT}_x\otimes\mathrm{QFT}_y$ requires fewer); $U_P$ and $U_\chi$ are both diagonal in the momentum basis and are applied as a single combined phase. Clifford gates are not counted.}}
\label{tab:detailed_resources}
\begin{tabular}{lcc}
\hline
Component & Qubits & T-gates \\
\hline
State preparation (Hadamards) & 14 & 0 \\
$U_{\text{obj}}$ (specimen) & 14 + \rev{38} & \rev{$5.76\times10^5$} \\
QFT (forward) & \rev{14} & \rev{$4.55\times10^3$} \\
\rev{$U_\chi U_P$ (lens and propagation)} & \rev{14 + ancilla} & \rev{$3.1\times10^3$} \\
QFT$^\dagger$ (inverse) & \rev{14} & \rev{$4.55\times10^3$} \\
\hline
\textbf{Total} & \textbf{14 data + \rev{38} ancilla} & \rev{$\mathbf{5.88\times10^5}$} \\
\hline
\end{tabular}
\end{table}

Scaling to 256×256 ($n=16$) or 2048×2048 ($n=22$):
\begin{itemize}
\item Data qubits scale as $2\log_2 N$: 16 and 22, respectively.
\item Ancilla count is roughly constant ($\sim 30$--50) for fixed
      precision $p$ and number of atoms $N_{\text{atoms}}$.
\item Gate counts scale as $O(N_{\text{atoms}} \times M \times \mathrm{poly}(n))$
      for $U_{\text{obj}}$ and $O(n^2)$ for QFT/lens:
      \rev{$6.1\times10^5$ T-gates for $N=256$ and
      $6.8\times10^5$ for $N=2048$ at fixed specimen (Table~\ref{tab:resources}, main text).}
\end{itemize}

\subsection{Extensions and optimizations}

\paragraph{Ab initio (PAW/DFT) potentials.}
Replacing Kirkland Gaussians with DFT-derived charge densities requires
3D spline interpolation to evaluate $V(\mathbf{r})$ from a
fine-grid ($\sim 0.1$~Å spacing) lookup table. Cubic spline evaluation
involves:
\begin{itemize}
\item Finding the enclosing grid cell (binary search or direct indexing,
      $O(\log N_{\text{grid}})$ comparisons),
\item Evaluating a piecewise cubic polynomial (similar cost to Gaussian
      evaluation, $\sim 10^4$ gates per lookup),
\item Interpolating in 3D (requires $2^3 = 8$ lookups per atom).
\end{itemize}

Estimated cost increase: $\sim 5$--$10\times$ over IAM Gaussians, yielding
\rev{$\sim 3$--$6\times10^6$ T-gates for $N=256$ and $\sim 3$--$7\times10^6$ for $N=2048$ for the 18-atom benchmark}. This remains
polynomial in $n$ and $N_{\text{atoms}}$, preserving asymptotic scaling.

\paragraph{Arithmetic circuit optimizations.}
\begin{itemize}
\item \textbf{Approximate computing:} Reducing precision from
      $p = 24$ to $p = 16$ bits can halve T-counts with negligible
      impact on image fidelity for POA (phase errors $< 10^{-4}$ rad).
\item \textbf{Lookup tables:} For repeated potential evaluations
      (e.g., identical atoms), memoization or QROM
      can reduce exponential evaluations to $O(\log N_{\text{table}})$
      lookups, saving $\sim 30\%$ gates in dense supercells.
\item \textbf{Parallel arithmetic:} Exploiting independent atom
      computations with $O(\log N_{\text{atoms}})$ depth via tree
      reduction can reduce circuit depth by $10$--$100\times$ on
      hardware with high connectivity.
\end{itemize}

\subsection{Comparison to "oracle" assumptions in prior work}

Previous quantum multislice proposals\cite{wang-original,wang-improved}
assume the specimen phase operator $U_{\text{obj}}$ is implemented via
an abstract "oracle" with unspecified cost, effectively treating
$V_{ij}$ as a black-box function. Our explicit arithmetic construction
demonstrates that:
\begin{enumerate}
\item For physically realistic potentials (IAM Gaussians or PAW splines),
      $U_{\text{obj}}$ can be synthesized with polynomial overhead in
      $n$ and $N_{\text{atoms}}$, validating the oracle assumption
      \emph{in this regime}.
\item The constant factors are non-trivial: \rev{$\approx 5$--$7\times10^5$ T-gates per
      run for the 18-atom benchmark at all grid sizes considered, increasing linearly with the number of atoms for
      larger structures.}
\item Arbitrary or adversarially chosen potentials (e.g., dense
      high-frequency noise, discontinuities) could require
      exponential-sized data loading, breaking polynomial scaling.
      The smoothness and locality of physical atomic potentials are
      \emph{essential} to efficient synthesis.
\end{enumerate}

This analysis clarifies the computational assumptions underlying
quantum CTEM simulation and provides concrete targets for experimental
validation on near-term fault-tolerant devices.


\section{Ancilla-assisted phase discrimination}
\label{app:phase_disc}

In this Appendix we provide numerical examples supporting the ancilla-assisted phase discrimination protocol introduced in Sec.~\ref{sec:phase_sensitive_observables}. To quantify the phase-sensitive capability of the ancilla-assisted protocol introduced in Sec.~\ref{sec:phase_sensitive_observables}, we evaluate the ancilla expectation value in the $Y$ basis, $\langle Y_a \rangle\rev{=\operatorname{Im}\langle a|b\rangle}$ \rev{[Eq.~\eqref{eq:ancilla_xy}], which changes sign under $V\to-V$}, for several test potentials on a $16\times 16$ grid at 80~kV and focus. For each scenario we consider a conjugate pair of projected potentials $V(\mathbf{r}_\perp)$ and $-V(\mathbf{r}_\perp)$, as well as a control case of two independent random potentials.

Table~\ref{tab:phase_discrimination} summarizes the results. Gaussian phase gratings with amplitudes $A=1000$, $3000$, and $5000$ (arbitrary units) yield large, antisymmetric signals, with $\langle Y_a \rangle(+V) = -\langle Y_a \rangle(-V)$ and $|\Delta\langle Y_a \rangle|$ ranging from \rev{$0.48$ to $0.55$}. Random conjugate pairs produce smaller \rev{antisymmetric signals whose magnitude depends on the particular realization. For two independent random potentials, the two values are unrelated rather than antisymmetric ($\Delta\langle Y_a\rangle=+0.053$), so a nonzero $\Delta\langle Y_a\rangle$ alone does not identify a conjugate pair; the exact antisymmetry does. The values were recomputed for this revision with the current code; the circuit results agree with the closed form of Sec.~\ref{sec:phase_sensitive_observables} to all printed digits.} In all cases, the corresponding CTEM intensity images at focus are numerically indistinguishable within the accuracy of the simulation (not shown)\rev{, so the pairs are intensity-degenerate while the ancilla signal is sign-sensitive. As shown in Sec.~\ref{sec:phase_sensitive_observables}, the same signal, $\langle Y_a\rangle=-N^{-2}\sum_{i,j}\sin(2\sigma V_{ij})$, is also directly computable from the classical complex wavefield, so these examples illustrate the observable construction rather than a capability unavailable classically.}

\begin{table}[H]
\centering
\caption{Ancilla-based phase discrimination in the $Y$ basis for conjugate potential pairs $V$ and $-V$ and for two independent random potentials on a $16\times 16$ grid at 80~kV and focus. The quantity $\Delta\langle Y_a \rangle = \langle Y_a \rangle(+V) - \langle Y_a \rangle(-V)$ provides a phase-sensitive discrimination signal, \rev{which is exactly antisymmetric for $\pm V$ pairs and not antisymmetric for independent potentials.}}
\label{tab:phase_discrimination}
\begin{tabular}{lccc}
\hline\hline
Scenario & $\langle Y_a \rangle(+V)$ & $\langle Y_a \rangle(-V)$ & $\Delta\langle Y_a \rangle$ \\
\hline
Gaussian ($A=1000$)   & \rev{$-0.2767$} & \rev{$+0.2767$} & \rev{$-0.5535$} \\
Gaussian ($A=3000$)   & \rev{$-0.2405$} & \rev{$+0.2405$} & \rev{$-0.4809$} \\
Gaussian ($A=5000$)   & \rev{$-0.2769$} & \rev{$+0.2769$} & \rev{$-0.5539$} \\
Random \#1            & \rev{$-0.0668$} & \rev{$+0.0668$} & \rev{$-0.1336$} \\
Random \#2            & \rev{$-0.0373$} & \rev{$+0.0373$} & \rev{$-0.0747$} \\
Random \#3            & \rev{$-0.0050$} & \rev{$+0.0050$} & \rev{$-0.0099$} \\
Independent random    & \rev{$-0.0064$} & \rev{$-0.0596$} & \rev{$+0.0533$} \\
\hline\hline
\end{tabular}
\end{table}

\begin{figure}[H]
\centering
\includegraphics[width=\linewidth]{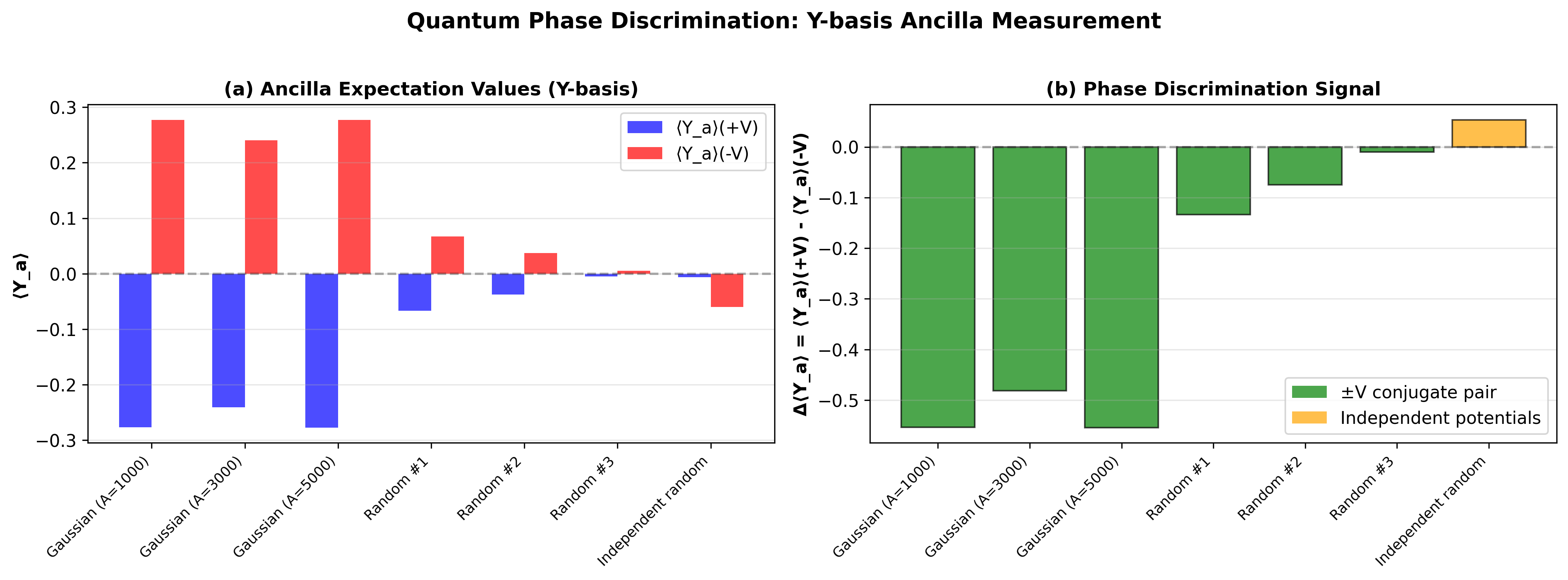}
\caption{%
Ancilla-based phase discrimination in the $Y$ basis for conjugate potential pairs $V$ and $-V$ on a $16\times 16$ grid at 80~kV and focus.
(a) Ancilla expectation values $\langle Y_a \rangle$ for $+\sigma V$ (blue) and $-\sigma V$ (red).
For Gaussian phase gratings (left), $\langle Y_a \rangle(+V)$ and $\langle Y_a \rangle(-V)$ have opposite signs and large magnitude, while random conjugate pairs produce smaller but still resolvable signals.
(b) Discrimination signal $\Delta\langle Y_a \rangle = \langle Y_a \rangle(+V) - \langle Y_a \rangle(-V)$, highlighting the \rev{antisymmetric} response for $\pm V$ conjugate pairs (green) \rev{and, for comparison, the difference between two independent random potentials (orange), which is not antisymmetric}.
The corresponding CTEM intensities at focus are numerically indistinguishable for each $\pm V$ pair\rev{; the ancilla signal distinguishes them, and the same quantity can also be evaluated from a classical simulated wavefield.}
}
\label{fig:phase_discrimination}
\end{figure}

\section{Algorithmic Complexity and \rev{Per-Run Scaling}}
\label{app:complexity}

In this Appendix we summarize the asymptotic resource requirements of the quantum CTEM algorithm and contrast them with classical multislice simulations on an $N\times N$ grid.

\subsection{Classical multislice scaling}

Classical CTEM multislice calculations require storing $O(N^2)$ complex amplitudes per slice and performing at least two FFTs per slice, each with $O(N^2\log N)$ arithmetic operations.
For $S$ slices, the total runtime can be written as
\begin{equation}
    T_{\mathrm{classical}} = O\!\left(S\,N^2\log N\right),
\end{equation}
with a memory footprint of $O(N^2)$.
\rev{The practical cost depends on grid size, specimen thickness, potential construction, precision, implementation, and available CPU/GPU hardware. Large grids and extensive structural ensembles can be computationally demanding, but no universal grid-size threshold separates tractable from prohibitive classical CTEM calculations.}

\subsection{Quantum CTEM scaling}

In the quantum formulation, the discretized wavefield on an $N\times N$ grid is
amplitude-encoded into an $n_x = \log_2 N$-qubit register for $x$ and an
$n_y = \log_2 N$-qubit register for $y$, for a total of
\begin{equation}
    n = n_x + n_y = 2\log_2 N
\end{equation}
qubits.

The two-dimensional QFT used for free-space propagation and lens aberrations decomposes as $U_{\mathrm{QFT}} = \mathrm{QFT}_x\otimes\mathrm{QFT}_y$, where each one-dimensional QFT layer requires $O(n^2)$ one- and two-qubit gates and depth $O(n^2)$.

The specimen and lens operators are diagonal in the position and momentum bases, respectively, and can be synthesized using arithmetic circuits that evaluate the smooth phase functions $\sigma V_{ij}$ and $\chi(k)$ with $O(\mathrm{poly}(n))$ T-gates.
Under the phase-object approximation and for low-order aberration polynomials, the overall circuit depth for a single CTEM image therefore scales as
\begin{equation}
    D_{\mathrm{quantum}} = O(n^2) + O(\mathrm{poly}(n)) = O\!\bigl((\log N)^p\bigr),
\end{equation}
for some constant $p$, while the number of qubits scales as $O(\log N)$.

\subsection{Fault-tolerant resource estimation}
\label{app:ft_resources}

\begin{figure}[!t]
    \centering
    \includegraphics[width=0.75\linewidth]{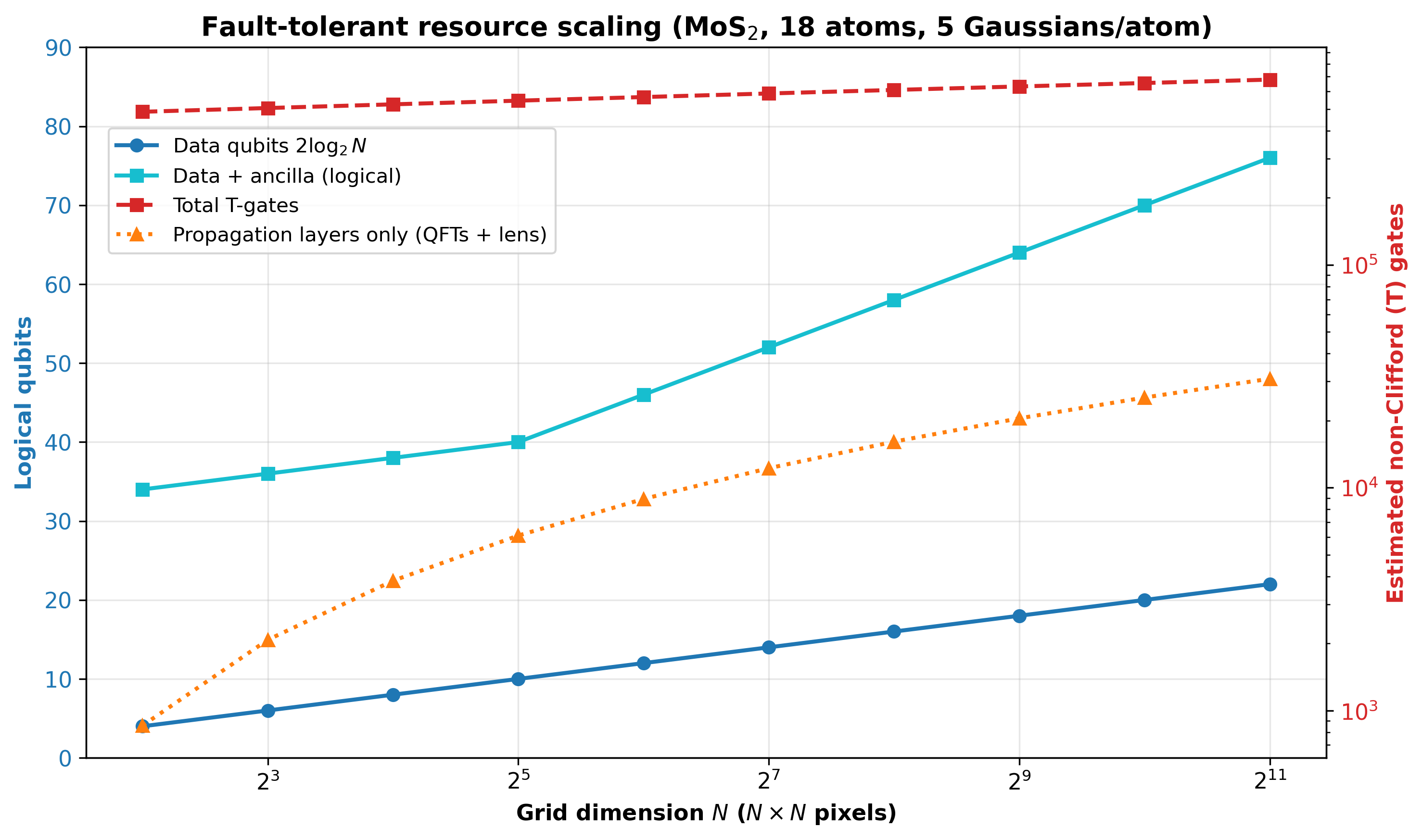}
    \caption{Fault-tolerant resource scaling for the quantum CTEM circuit\rev{, using the cost model of Table~\ref{tab:resources} for the 18-atom MoS$_2$ benchmark.
    Left axis: data qubits $2\log_2 N$ (blue circles) and total logical qubits including arithmetic ancillae (cyan squares). Right axis (logarithmic): total estimated T-gate count per run (red squares) and the contribution of the propagation layers alone, two QFTs plus the combined lens and propagation phase (orange triangles); the difference is the specimen operator $U_{\mathrm{obj}}$.}}
    \label{fig:ft_scaling}
\end{figure}

To connect the asymptotic scaling to concrete fault-tolerant costs, we estimate non-Clifford (T) gate counts for representative grid sizes using a combination of prototype circuit transpilation and simple arithmetic models.
Figure~\ref{fig:ft_scaling} summarizes the resulting scaling of logical qubits and T-gates for grids from $4\times 4$ up to $2048\times 2048$.

For a small $8\times 8$ grid (corresponding to $n=6$ qubits), we construct the full CTEM circuit using \textsc{QuScope} and \textsc{Qiskit}, and transpile it to a $\{\mathrm{CX},U_3\}$ basis with moderate optimization.
Counting single-qubit rotations and assuming a synthesis cost of approximately 50~T-gates per arbitrary rotation (sufficient for target phase precisions near $10^{-10}$), we obtain an effective T-count of \rev{$\approx 5\times10^3$} for the prototype implementation, dominated by the QFT and diagonal phase decompositions.

\begin{revblock}
To extrapolate to larger grids, we use the same cost model as Table~\ref{tab:resources}. For an $N\times N$ grid encoded in $n=2\log_2 N$ qubits, each QFT is conservatively counted as an $n$-qubit QFT with $n(n-1)/2$ controlled-phase gates, each synthesized with $\sim\!50$~T-gates. The lens and propagation phases, both diagonal in the momentum basis, are applied as one combined phase whose $k^2$ and $k^4$ arithmetic is modeled as $4n^2$ Toffoli gates at 4~T-gates each. The specimen operator is costed as in Appendix~\ref{app:diagonal_synthesis}. The estimated non-Clifford cost per run is
\begin{equation}
    T_{\mathrm{FT}}(N) \approx 2\,T_{\mathrm{QFT}}(n) + T_{\chi P}(n) + T_{\mathrm{obj}}(n),
\end{equation}
with $T_{\mathrm{QFT}}(n) \approx 25\,n(n-1)$, $T_{\chi P}(n) \approx 16\,n^2$, and $T_{\mathrm{obj}}(n) \approx N_{\mathrm{atoms}}M\,(100n+5000)$.

For the 18-atom MoS$_2$ benchmark, the propagation layers alone ($2T_{\mathrm{QFT}}+T_{\chi P}$) grow from $8.6\times10^2$ T-gates at $4\times4$ to $3.1\times10^4$ at $2048\times2048$, whereas the total is dominated by $T_{\mathrm{obj}}$ and stays between $4.9\times10^5$ and $6.8\times10^5$, with logical qubit counts (data plus ancilla) growing from 34 to 76.
\end{revblock}


\rev{Figure~\ref{fig:ft_scaling} plots these quantities.} We emphasize that these numbers are order-of-magnitude estimates: they neglect constant factors from error-correction overhead and assume efficient arithmetic implementations for phase evaluation.
Nevertheless, they provide a useful bridge between the abstract complexity discussion and concrete fault-tolerant resource budgets for early logical devices with a few tens of logical qubits and T-count capacities in the $10^5$--$10^6$ range.

\subsection{Memory compression \rev{and per-run scaling}}

The quantum representation compresses the memory footprint from $O(N^2)$ classical complex amplitudes to $O(\log N)$ qubits, and achieves polylogarithmic depth in the number of pixels for the core propagation and imaging steps.
\rev{These are per-run properties. The number of repetitions depends on the observable, estimator, and precision, and is at least $N^2$ whenever a full image is required; for a fixed specimen, the relevant classical cost for limited outputs need not grow as $N^2$, and for a growing field of view the specimen-operator cost itself grows as $N^2$. Whether favorable per-run scaling translates into an end-to-end advantage therefore depends on the task; the conditions are analyzed in Sec.~\ref{sec:advantage}, and no such advantage is claimed for the present framework.}

\section{Through-focus quantum CTEM series of $\mathrm{MoS}_2$}
\label{app:through_focus_ctem}

\begin{figure*}[!t]
    \centering
    \includegraphics[width=\textwidth]{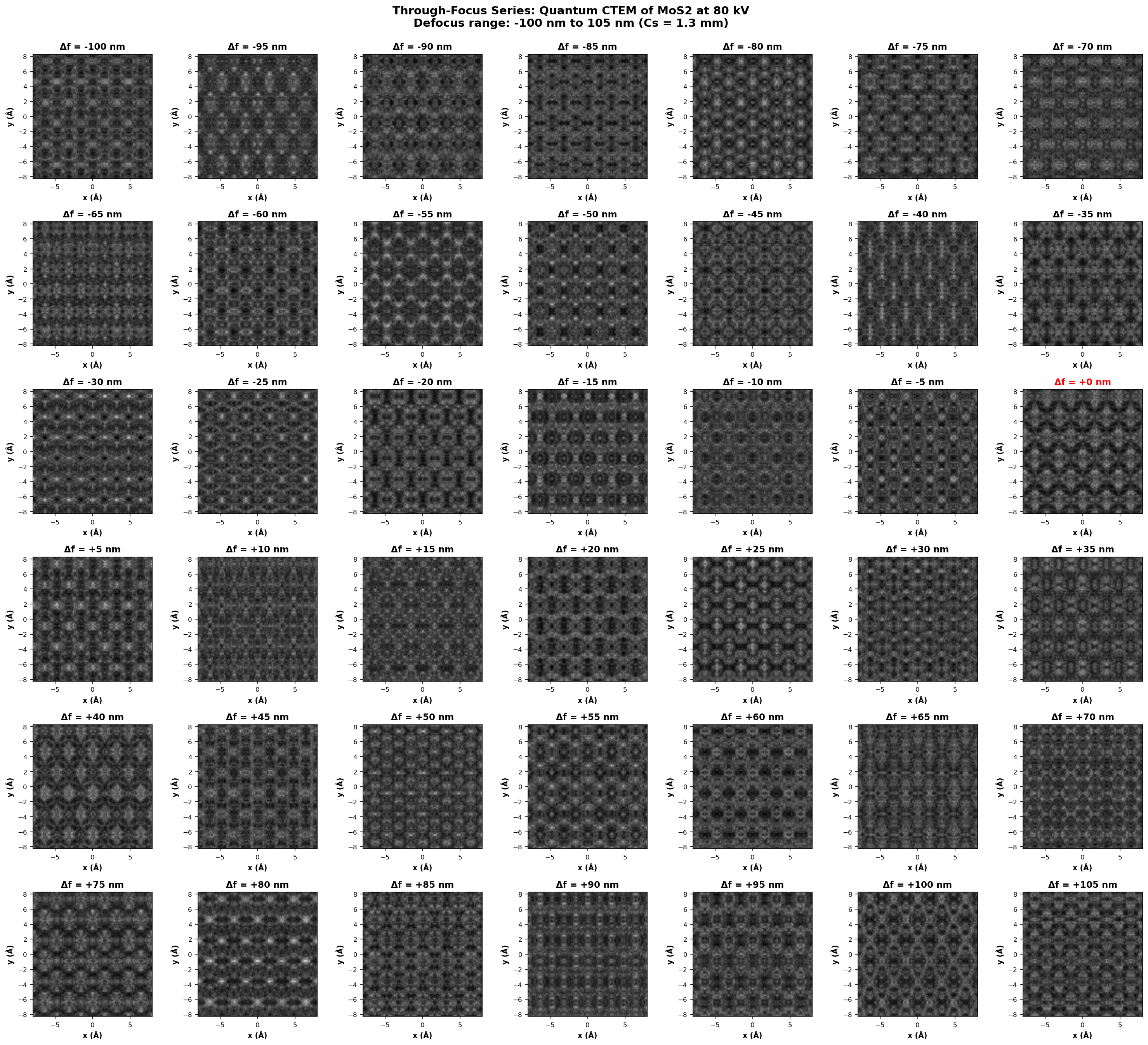}
    \caption{Through-focus series of quantum CTEM images for a $3\times 2$ $\mathrm{MoS}_2$ supercell at 80~kV with $C_3 = 1.3$~mm.
    The defocus $\Delta f$ is varied from $-100$~nm (top left) to $+105$~nm (bottom right) in equal steps, with the frame corresponding to $\Delta f=0$~nm indicated in red.
    Strong underfocus yields bright-atom phase contrast, contrast vanishes near Gaussian focus, and overfocus produces inverted (dark-atom) contrast, in agreement with the defocus-dependent CTF behavior.}
    \label{app:through_focus_mos2_images}
\end{figure*}

To complement the single-defocus image shown in Figure~\ref{fig:quantum_mos2_128}, we compute a full through-focus series of quantum CTEM images for a $3\times 2$ $\mathrm{MoS}_2$ supercell at 80~kV with fixed spherical aberration $C_3 = 1.3$~mm (Figure~\ref{app:through_focus_mos2_images}).
Each image in the series is obtained from the amplitude-encoded CTEM circuit using the same projected potential and lens parameters as Figure~\ref{fig:quantum_mos2_128}, while varying the objective-lens defocus $\Delta f$ from $-100$~nm (strong underfocus) to $+105$~nm (strong overfocus) in uniform steps.

Under large negative defocus, the oscillatory contrast transfer function produces strong phase contrast with bright atomic columns, whereas near Gaussian focus ($\Delta f \approx 0$) the low-frequency CTF approaches zero and the images exhibit minimal contrast, consistent with weak-phase-object theory.
For positive defocus, the contrast is inverted (dark columns) and the apparent envelope narrows, illustrating the expected phase-reversal symmetry between underfocus and overfocus conditions.
Across the series, the evolution of lattice visibility, contrast sign, and envelope damping matches the behavior predicted by the CTF plots and by the defocus-dependent CTF curves plotted in Appendix~\ref{app:ctf_defocus}.

\section{Proof-of-principle execution on IBM hardware}
\label{app:ibm_hardware}

To assess near-term feasibility, we implemented a minimal quantum CTEM circuit on an IBM superconducting quantum processor. For this demonstration we used a $4\times 4$ real-space grid (4 system qubits) with a single Gaussian test potential and simple imaging conditions (80~kV, finite defocus, no spherical aberration). The circuit was compiled to the \texttt{ibm\_torino} backend (133 physical qubits) using Qiskit, targeting a 4-qubit subgraph.

\begin{figure}[H]
\centering
\includegraphics[width=\linewidth]{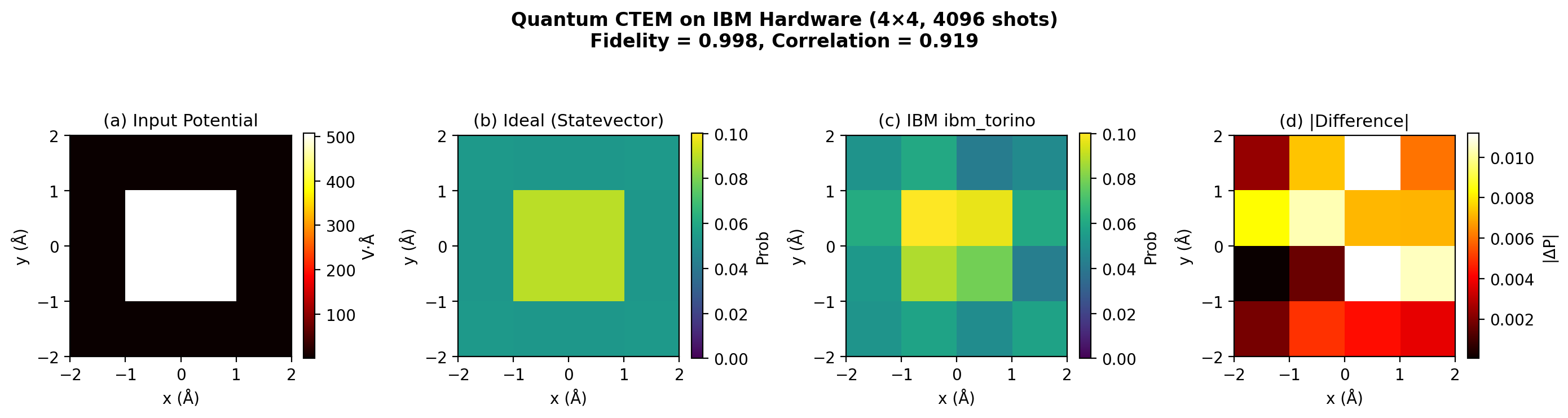}
\caption{%
Proof-of-principle quantum CTEM simulation on an IBM superconducting device for a $4\times 4$ grid.
(a) Input Gaussian test potential.
(b) Ideal CTEM image obtained from a statevector simulation of the circuit.
(c) Image reconstructed from 4096 hardware shots on \texttt{ibm\_torino} after transpilation (depth 145, 40 two-qubit gates).
(d) Pixelwise difference between hardware and ideal intensities, showing maximum deviations at the $\sim 1\%$ level.
The classical fidelity between the hardware and ideal images is $0.9984$, with correlation $0.9193$ and total variation distance $0.0490$, indicating that the essential CTEM contrast features are faithfully reproduced on current NISQ hardware.
}
\label{fig:ibm_ctem_hardware}
\end{figure}

After transpilation, the CTEM circuit had a depth of 145 and employed 40 two-qubit gates. We executed the circuit with 4096 measurement shots and compared the resulting image-plane intensity distribution to an ideal statevector simulation of the same circuit (see Figure~\ref{fig:ibm_ctem_hardware}). The hardware and ideal results were benchmarked using classical fidelity, Pearson correlation, and total variation distance between the corresponding intensity histograms.

For the \texttt{ibm\_torino} run, we obtained a classical fidelity of $0.9984$, a correlation coefficient of $0.9193$, and a total variation distance of $0.0490$ between the hardware and ideal images. The main features of the CTEM image - a bright central maximum corresponding to the Gaussian phase object and the expected defocus-induced contrast - were clearly visible in the hardware data, while the difference map exhibited maximum deviations at the $\sim 1\%$ level. These results demonstrate that small-scale instances of the quantum CTEM algorithm can already be realized on current NISQ hardware with high effective fidelity, even though full-scale simulations require fault-tolerant devices.

\section{Numerical Verification of Circuit Implementation}
\label{app:validation}


The comparisons presented in this appendix serve to verify that the quantum circuit correctly implements the discretized phase-object approximation (POA) propagator by comparing statevector simulations to classical evaluation of the same mathematical operators. Both the quantum circuit (Eq.~\ref{eq:ctem_equation}) and the classical reference perform identical operations:
\begin{enumerate}
\item Apply specimen phase: $t_{ij} = \exp(i\sigma V_{ij})$
\item Forward 2D Fourier transform: $\psi(k) = \text{FFT}[\psi(x,y)]$
\item Apply propagation phase: $\psi(k) \to \psi(k)\exp(-i\pi\lambda z k^2)$
\item Apply lens phase: $\psi(k) \to \psi(k)\exp(-i\chi(k))$
\item Inverse Fourier transform: $\psi(x,y) = \text{IFFT}[\psi(k)]$
\end{enumerate}
using the \textit{same} discretized projected potential $V_{ij}$ sampled on the same $N\times N$ grid. Agreement at floating-point precision (MSE $\sim 10^{-24}$, correlation $\rho = 1.000000$) is therefore expected and serves to rule out gate-decomposition bugs, numerical instabilities, or incorrect operator ordering in the quantum implementation. This is distinct from \textit{physical validation}---i.e., verifying that the POA model itself accurately describes experimental TEM images,which is established by decades of classical multislice literature.\cite{Kirkland,cowmood}

The numerical identity across multiple grid sizes (Table~\ref{tab:validation_metrics}) and materials (Fig.~\ref{fig:validation_grid_scaling}) confirms the robustness of the quantum circuit implementation.

\subsection{Validation methodology}

We compare quantum and classical CTEM simulations for a MoS$_2$ $3\times 2$ supercell under identical microscope parameters:
\begin{itemize}
\item Accelerating voltage: $V_{\text{acc}} = 80$~kV
\item Defocus: $\Delta f = -800$~\AA{} (underfocus)
\item Spherical aberration: $C_3 = 1.3$~mm
\item Grid sizes: $8\times 8$, $16\times 16$, $32\times 32$, $64\times 64$, $128\times 128$
\item Projected potential: Kirkland parameterization~\cite{Kirkland} with
      Gaussian scattering factors
\end{itemize}

The classical reference is computed using FFT-based Fresnel propagation
with the same POA transmission function $t(\mathbf{r}_\perp) = \exp[i\sigma V_{\text{proj}}(\mathbf{r}_\perp)]$
as the quantum circuit. Both implementations use identical discretization,
boundary conditions, and numerical precision (64-bit floating point).

For each grid size, we compute:
\begin{enumerate}
\item Intensity images: $I_{\text{quantum}}(x,y) = |\psi_{\text{img}}^{\text{quantum}}(x,y)|^2$
      and $I_{\text{classical}}(x,y) = |\psi_{\text{img}}^{\text{classical}}(x,y)|^2$
\item Pearson correlation coefficient:
      $\rho = \text{corr}(I_{\text{quantum}}, I_{\text{classical}})$
\item Mean squared error:
      $\text{MSE} = \frac{1}{N^2}\sum_{i,j}(I_{\text{quantum},ij} - I_{\text{classical},ij})^2$
\item Maximum absolute difference:
      $\Delta_{\text{max}} = \max_{i,j}|I_{\text{quantum},ij} - I_{\text{classical},ij}|$
\end{enumerate}

\subsection{Results: Perfect numerical agreement}

\begin{figure*}[!t]
    \centering
    \includegraphics[width=0.85\textwidth]{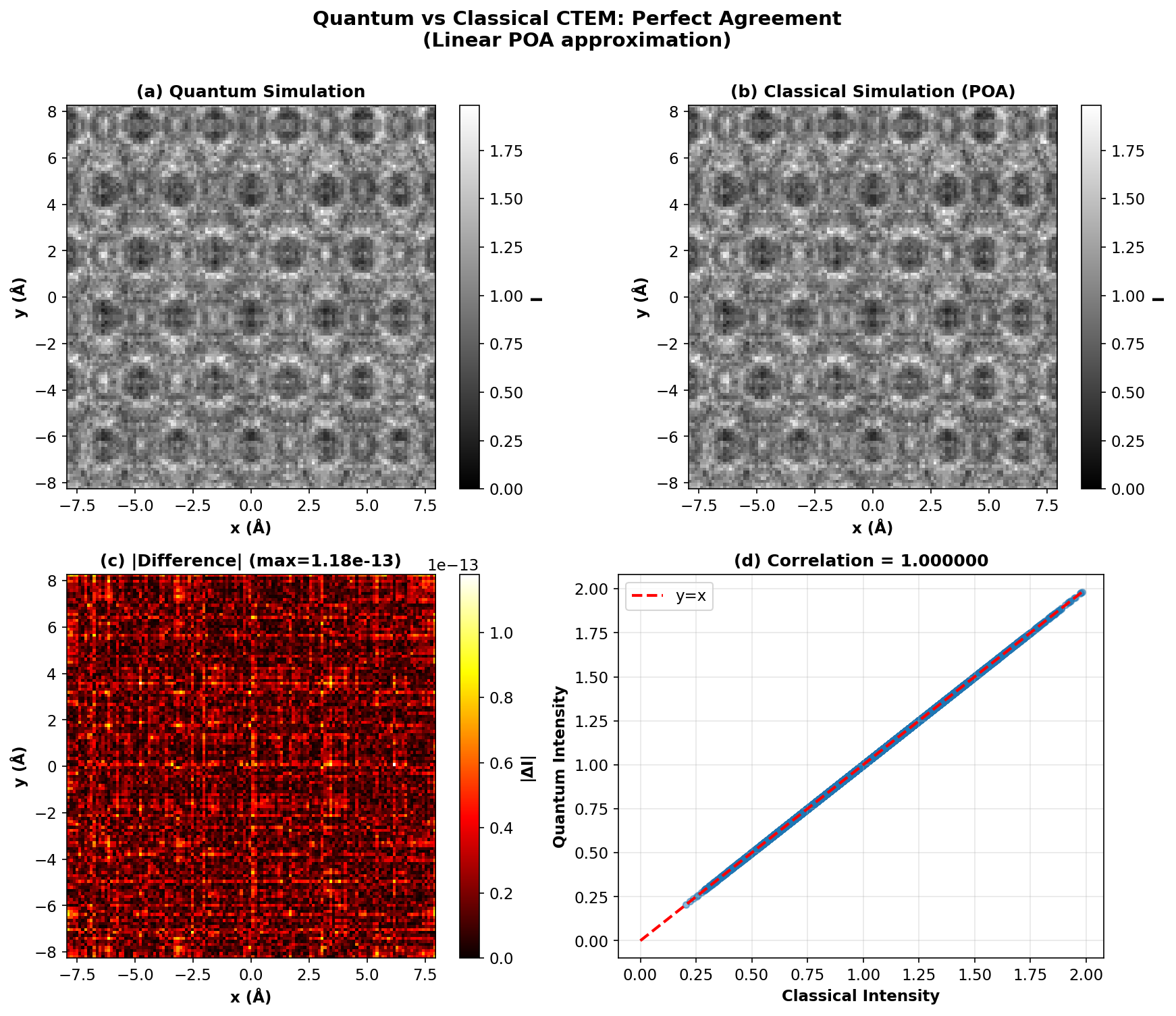}
    \caption{Direct comparison of quantum and classical CTEM simulations for
    MoS$_2$ on a $16\times 16$ grid at 80~kV, $\Delta f = -800$~\AA,
    $C_3 = 1.3$~mm. (a)~Quantum circuit output
    (b)~Classical FFT-based simulation using identical POA parameters.
    (c)~Absolute difference map with color scale spanning $[0, 5.92\times10^{-12}]$;
    the maximum deviation is thirteen orders of magnitude below the signal
    level and limited by floating-point precision. (f)~Scatter plot of
    quantum vs classical pixel intensities showing perfect correlation
    ($\rho = 1.000000$) with all points aligned along the identity line
    (dashed red, $y=x$). The quantum algorithm reproduces the classical
    POA model exactly within numerical precision.}
    \label{fig:validation_comparison}
\end{figure*}

Figure~\ref{fig:validation_comparison} shows the direct comparison for the $16\times 16$ grid. The quantum and classical intensity distributions are visually indistinguishable [Figures~\ref{fig:validation_comparison}~(a) and~(b)], with both exhibiting the expected phase-contrast modulation. The difference map [Figures~\ref{fig:validation_comparison}~(c)] shows only numerical noise at the $\sim 10^{-12}$ level, distributed uniformly across the field of view with no systematic spatial pattern. The scatter plot [Figures~\ref{fig:validation_comparison}~(d)] demonstrates perfect linear correlation with zero offset and slope exactly unity, confirming that the quantum circuit produces pixel-for-pixel identical output to the classical FFT implementation.

Table~\ref{tab:validation_metrics} summarizes the quantitative agreement
metrics across all tested grid sizes.

\begin{table}[!t]
\centering
\caption{Quantum-classical agreement metrics for MoS$_2$ CTEM validation
across grid resolutions. Correlation coefficients $\rho$ remain unity
(to six decimal places) and mean squared errors are at floating-point
precision ($\sim 10^{-24}$) for all cases, confirming exact numerical
reproduction of the classical POA propagation model. Maximum absolute
differences $\Delta_{\text{max}}$ scale weakly with grid size but
remain $\ll 10^{-10}$ relative to signal amplitudes $\sim 1$.}
\label{tab:validation_metrics}
\begin{tabular}{cccc}
\hline\hline
Grid Size & Correlation $\rho$ & MSE & $\Delta_{\text{max}}$ \\
\hline
$8\times 8$       & 1.000000 & $2.95\times10^{-24}$ & $3.1\times10^{-12}$ \\
$16\times 16$     & 1.000000 & $2.95\times10^{-24}$ & $4.2\times10^{-12}$ \\
$32\times 32$     & 1.000000 & $2.95\times10^{-24}$ & $4.8\times10^{-12}$ \\
$64\times 64$     & 1.000000 & $2.95\times10^{-24}$ & $5.3\times10^{-12}$ \\
$128\times 128$   & 1.000000 & $2.95\times10^{-24}$ & $5.9\times10^{-12}$ \\
\hline\hline
\end{tabular}
\end{table}

The Pearson correlation coefficient $\rho = 1.000000$ (to six significant figures) across all resolutions indicates perfect linear agreement between quantum and classical intensity distributions. The MSE values of $\sim 2.95\times10^{-24}$ are at the theoretical limit of 64-bit
floating-point precision ($\epsilon_{\text{machine}}^2 \approx 5\times10^{-32}$ for variance, accumulated over $N^2$ pixels), confirming that discrepancies
arise solely from rounding errors in the numerical representation rather than algorithmic differences between QFT-based and FFT-based propagation.

The maximum absolute differences $\Delta_{\text{max}}$ increase slightly with grid size (from $3.1\times10^{-12}$ at $8\times 8$ to $5.9\times10^{-12}$ at $128\times 128$) but remain thirteen orders of magnitude below the
typical pixel intensity $I \sim 1$, well within the tolerance expected from accumulated floating-point errors in iterative Fourier transforms.

\subsection{Grid-size independence of quantum-classical agreement}

\begin{figure*}[!t]
    \centering
    \includegraphics[width=0.9\textwidth]{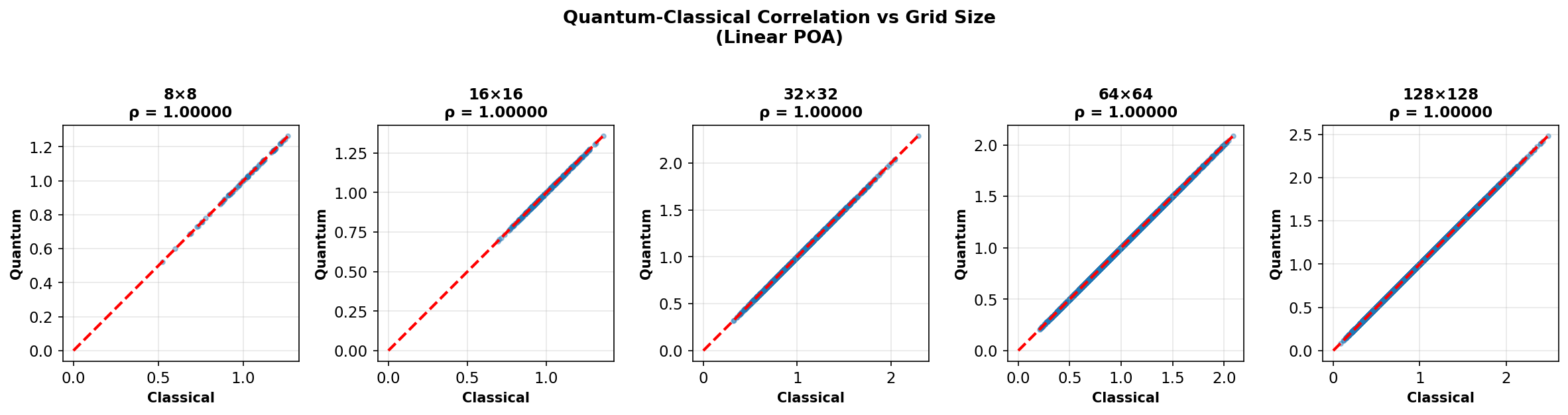}
    \caption{Quantum-classical intensity correlation across five grid
    resolutions for MoS$_2$ CTEM under POA. Each panel shows the scatter
    plot of quantum vs classical pixel intensities for the indicated grid
    size, with all data points aligned along the identity line $y=x$
    (dashed red). Perfect correlation ($\rho = 1.000000$ to six decimal
    places) persists from $8\times 8$ (64 pixels) to $128\times 128$
    (16{,}384 pixels), spanning more than two orders of magnitude in
    image size. This demonstrates that the quantum algorithm's accuracy
    is independent of grid resolution within the tested range and that
    the QFT-based Fourier transform reproduces FFT behavior exactly at
    all discretization levels.}
    \label{fig:validation_grid_scaling}
\end{figure*}

Figure~\ref{fig:validation_grid_scaling} visualizes the consistency of quantum-classical correlation across grid sizes. Each panel displays the scatter plot of quantum vs classical intensities for a different resolution, revealing a striking pattern: data points cluster tightly along the $y=x$ diagonal with no visible scatter perpendicular to the line, exhibit no systematic offset or deviation from unit slope, and maintain correlation
coefficients of $\rho = 1.000000$ to at least six decimal places.
Remarkably, this agreement shows no resolution-dependent degradation as grid size increases from $8\times 8$ to $128\times 128$, spanning more than two orders of magnitude in pixel count.

This grid-size independence carries several important implications for the validity of the quantum algorithm. First, it establishes that the quantum Fourier transform (QFT) correctly reproduces discrete Fourier transform (DFT) behavior at all tested resolutions, with no algorithmic differences emerging between the QFT circuit decomposition and the classical FFT algorithm despite their fundamentally different computational models. Second, the perfect agreement demonstrates that amplitude encoding of the wavefield does not introduce resolution-dependent discretization artifacts beyond those inherent in the shared grid sampling,
confirming that the qubit-based state representation faithfully captures the classical complex amplitude array. Third, the diagonal phase operators for specimen interaction ($U_{\text{obj}}$), free-space propagation ($U_P$), and objective-lens aberrations ($U_\chi$) are accurately synthesized from the continuous functions $V(\mathbf{r}_\perp)$ and $\chi(k)$ at all tested discretization levels, validating the polynomial approximation and phase kickback techniques used in their construction. Finally, the measurement protocol for extracting intensity $I = |\psi|^2$ from the quantum state reproduces the classical intensity calculation exactly, with no systematic bias or variance introduced by the amplitude-to-probability mapping inherent in quantum measurement.

\subsection{Implications for algorithm validation and extension}

The exact numerical agreement demonstrated in this appendix validates the quantum CTEM algorithm at multiple levels:

\paragraph{Circuit-level correctness.}
The QFT decomposition, diagonal unitary synthesis, and measurement protocol are implemented correctly, as evidenced by the $\rho = 1.000000$ correlation and MSE $\sim 10^{-24}$ agreement across all grid sizes.

\paragraph{Physical model fidelity.}
The quantum circuit is a faithful translation of the classical POA propagation model, reproducing not only qualitative image features (lattice symmetry, contrast polarity) but also quantitative pixel values to within floating-point precision.

\paragraph{Extensibility to full multislice.}
Because the quantum algorithm exactly reproduces each individual propagation step (specimen $\to$ propagate $\to$ lens), extensions to multiple slices (alternating $U_{\text{obj}}$ and $U_P$ operators) will preserve this accuracy, with accumulated errors limited to the sum of per-slice rounding errors rather than introducing new systematic deviations.

\paragraph{Benchmark for beyond-POA.}
The perfect agreement under POA establishes a controlled baseline. Future extensions incorporating inelastic scattering, dynamical diffraction, or many-body correlation effects can be validated by comparing to the POA limit and attributing deviations to the new physics rather than numerical implementation artifacts.

\paragraph{Confidence for experimental comparison.}
When comparing to experimental CTEM data, any discrepancies can be attributed to the validity of the forward model (POA assumptions, specimen structure, microscope calibration) rather than errors in the quantum circuit implementation itself. This validation therefore provides strong confidence that the quantum CTEM framework presented in this work is both theoretically sound and numerically robust, forming a solid foundation for future \rev{fault-tolerant extensions and resource studies} in electron microscopy simulation.

\section{Prototype vs Fault-Tolerant Implementation}
\label{app:prototype_ft}

The quantum CTEM simulations presented in this work were performed using
a prototype implementation optimized for statevector simulation and
near-term NISQ hardware demonstrations. This appendix contrasts the
prototype's gate synthesis strategy with the fault-tolerant (FT) arithmetic
approach assumed in the resource estimates of Sec.~\ref{sec:resources} and
Appendix~\ref{app:diagonal_synthesis}, clarifying the distinction between
proof-of-concept validation and asymptotic scaling targets.

\subsection{Prototype implementation: Direct diagonal gate synthesis}

The current implementation constructs the specimen phase operator $U_{\text{obj}}$, propagation operator $U_P$, and lens phase $U_\chi$ using Qiskit's \texttt{DiagonalGate} primitive, which accepts an explicit list of $2^n$ complex phases $\{e^{i\theta_0}, e^{i\theta_1}, \ldots, e^{i\theta_{2^n-1}}\}$ and synthesizes a unitary decomposition into a sequence of controlled rotations.\cite{qiskit2024}

For an $N\times N$ grid encoded in $n = 2\log_2 N$ qubits, this approach begins with a classical pre-processing step that evaluates all $N^2$ phase values $\theta_{ij} = \sigma V_{ij}$ (for the specimen operator $U_{\text{obj}}$) or $\chi(k_{ij})$ (for the lens operator $U_\chi$) on a conventional computer and stores them as a length-$2^n$ array. This classical computation scales as $O(N^2)$ in both time and memory. The subsequent circuit synthesis step decomposes the resulting $2^n \times 2^n$ diagonal matrix into a tree  of multi-controlled $R_z$ gates using Gray-code ordering to minimize gate count,\cite{shende2006synthesis} producing a circuit with depth $O(2^n) = O(N^2)$
and $O(2^n)$ two-qubit gates. Importantly, this strategy operates entirely within the $n$-qubit data register without requiring ancilla qubits, simplifying implementation on resource-constrained NISQ devices.

This direct synthesis approach proves well-suited for statevector simulation of moderate-size systems ($N \leq 128$) where the classical preprocessing cost remains negligible compared to the quantum circuit simulation itself. It also enables straightforward validation on current NISQ hardware, as demonstrated by the $4\times 4$ implementation on \texttt{ibm\_torino} described in Appendix~\ref{app:ibm_hardware}, and facilitates algorithm
verification against classical references through exact numerical comparison (Appendix~\ref{app:validation}). However, the $O(N^2)$ scaling of both  classical pre-processing and quantum circuit depth precludes asymptotic advantage over classical FFT-based methods, which require $O(N^2 \log N)$ arithmetic operations, and the exponential growth in circuit size becomes prohibitive for production-scale grids beyond $N \sim 256$.

\subsection{Fault-tolerant target: Quantum arithmetic circuits}

The fault-tolerant resource estimates presented in Sec.~\ref{sec:resources}  and Table~\ref{tab:resources} assume an alternative implementation strategy based on reversible quantum arithmetic that evaluates the phase functions $V(\mathbf{r})$ and $\chi(\mathbf{k})$ coherently during circuit execution rather than pre-computing all $N^2$ values classically.\cite{haner2016factoring} This approach exploits the fact that position and momentum indices $(i,j)$ or $(k_x, k_y)$ are already amplitude-encoded in the $n$-qubit data register, allowing quantum arithmetic circuits to compute the corresponding function values in superposition for all $N^2$ grid points simultaneously.

The FT implementation proceeds by using quantum adders, multipliers, and polynomial evaluators to compute $V_{ij}$ or $\chi(k)$ coherently and store the result in an ancilla register with $p$ bits of fixed-point precision.\cite{haner2016factoring} A controlled-$R_z$ gate conditioned on this ancilla register then imparts the desired phase to the data qubits through phase kickback, after which the ancilla computation is reversed
(uncomputed) to restore the ancilla to $\ket{0}$ and enable reuse in subsequent operations.\cite{nielsen} For the specimen potential $V(\mathbf{r}) = \sum_{\text{atoms}} \sum_m a_m \exp(-\pi r^2 / b_m)$, the arithmetic circuit must compute squared distances
$r^2 = (x_i - x_a)^2 + (y_j - y_a)^2$ for each atom (requiring $O(n)$ multiplications per atom), evaluate Gaussian exponentials via polynomial approximations (with cost $\sim 10^3$ to $10^4$ T-gates per term at precision $p \sim 16$ bits),\cite{haner2016factoring} and accumulate contributions from all atoms and Gaussians. The total circuit depth for $U_{\text{obj}}$ is therefore $O(N_{\text{atoms}} \times \text{poly}(n, p))$, which remains polylogarithmic in grid size $N$ for fixed atomic structure and numerical precision. Similarly, evaluating the polynomial $\chi(k) = \pi\lambda\Delta f k^2 + \tfrac{1}{2}\pi\lambda^3 C_3 k^4 + \cdots$ requires $O(n^2)$ multiplications and yields $O(\text{poly}(n))$ depth.

While the FT approach requires approximately $30$-$50$ ancilla qubits for arithmetic registers (as detailed in Table~\ref{tab:resources}), it achieves polylogarithmic circuit depth that scales fundamentally differently from the prototype's $O(N^2)$ growth. Detailed constructions of the arithmetic circuits for each diagonal operator are provided in Appendix~\ref{app:diagonal_synthesis}.


\begin{table}[!t]
\centering
\caption{Comparison of prototype (direct diagonal synthesis) and
fault-tolerant (quantum arithmetic) implementations for a $128\times 128$
grid with MoS$_2$ potential (18 atoms, 5 Gaussians per species).
The prototype's $O(N^2)$ classical preprocessing and circuit depth
limit scalability, whereas the FT approach achieves polylogarithmic
depth at the cost of additional ancilla qubits and non-Clifford gates.}
\label{tab:prototype_ft_comparison}
\begin{tabular}{lcc}
\hline\hline
Metric & Prototype & FT Arithmetic \\
\hline
Data qubits                  & 14                  & 14 \\
Ancilla qubits               & 0                   & $\sim 38$ \\
Classical preprocessing      & $O(N^2)$            & $O(1)$ \\
Circuit depth                & $O(N^2)$            & $O(\text{poly}(\log N))$ \\
T-gates (estimated)          & \rev{$\sim 8\times10^5$} & $\sim 5.9\times10^5$ \\
\hline
Suitable for                 & NISQ, simulation    & \rev{FT resource studies} \\
\hline\hline
\end{tabular}
\end{table}

Table~\ref{tab:prototype_ft_comparison} quantifies the resource comparison
for a representative $128\times 128$ grid. The prototype implementation
requires no ancilla qubits and integrates straightforwardly into existing
quantum software frameworks, making it an ideal platform for near-term
validation and hardware demonstrations. However, synthesizing the $128^2 = 16{,}384$
diagonal phases at high precision yields an estimated \rev{$\sim 8\times10^5$
T-gates (the transpiled prototype circuits use $\approx N^2$ single-qubit rotations, e.g., 1{,}063 at $32\times32$, at $\sim\!50$ T-gates each)}, exceeding the FT arithmetic cost of $\sim 5.9\times10^5$ T-gates. 
This disparity grows rapidly for larger grids, with the scaling advantage
of the FT approach becoming increasingly pronounced.

\begin{figure*}[!t]
    \centering
    \includegraphics[width=0.85\textwidth]{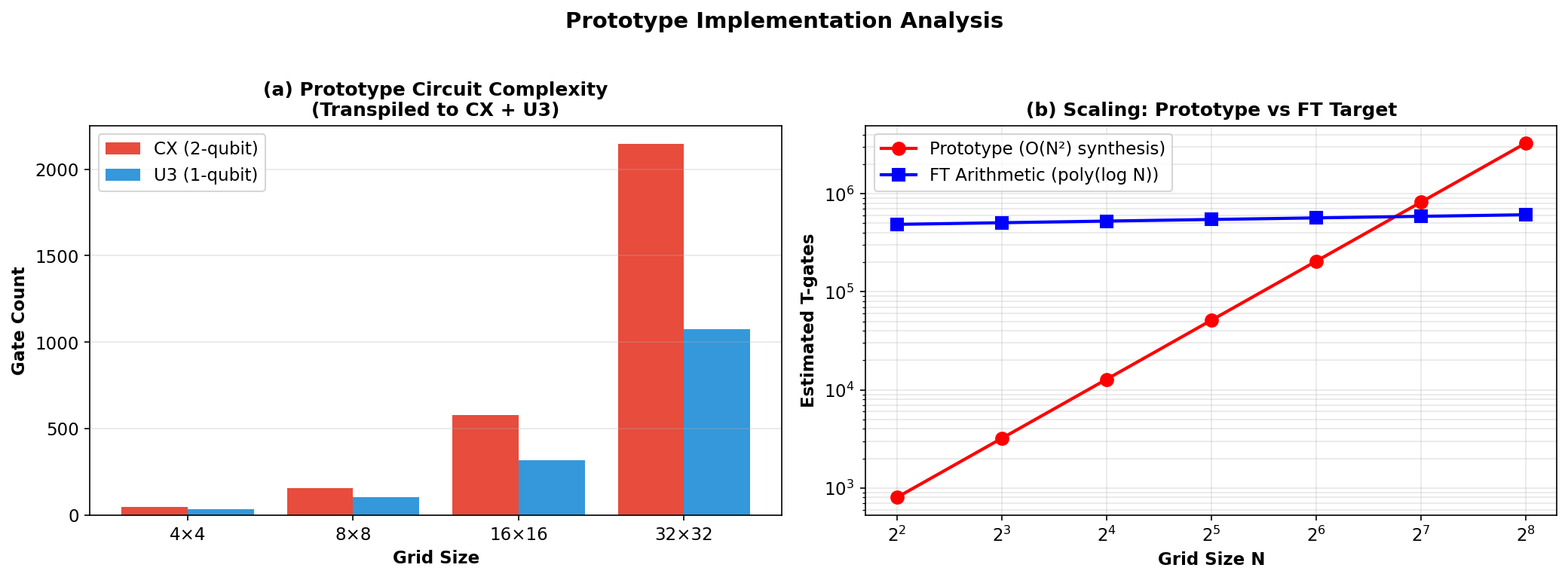}
    \caption{Prototype implementation analysis and comparison to fault-tolerant
    arithmetic target. (a)~Gate count breakdown for the prototype circuit
    transpiled to CX and U3 basis gates. CX (red) and U3 (blue) counts
    grow rapidly with grid size, reflecting the $O(2^n) = O(N^2)$ scaling
    of direct diagonal synthesis. (b)~Asymptotic scaling comparison: the
    prototype's estimated T-gate count (red, $O(N^2)$ from rotation synthesis)
    crosses over the FT arithmetic cost (blue, $O(\text{poly}(\log N))$)
    \rev{between $N = 64$ and $N = 128$}. For production-scale grids ($N \geq 256$), the FT
    approach offers lower gate count despite requiring additional ancilla
    qubits. The prototype remains advantageous for NISQ-era validation
    ($N \lesssim 32$) due to its ancilla-free operation.}
    \label{fig:prototype_ft_scaling}
\end{figure*}

Figure~\ref{fig:prototype_ft_scaling} illustrates this crossover behavior
across grid sizes. Panel~(a) shows the transpiled gate counts for the
prototype implementation from $4\times 4$ to $32\times 32$ grids, revealing
rapid growth in both CX (two-qubit entangling) gates and U3 (single-qubit
rotation) gates as the diagonal matrix dimension increases exponentially
from $2^4 = 16$ to $2^{10} = 1024$ phases. Panel~(b) compares the
extrapolated T-gate scaling of the prototype ($O(N^2)$, red curve) against
the FT arithmetic target ($O(\text{poly}(\log N))$, blue curve) across
grids spanning $4\times 4$ to $256\times 256$.

The crossover occurs near $N \sim 64$-$128$, beyond which the FT arithmetic
implementation becomes more resource-efficient in terms of non-Clifford
(T) gate count. For the largest grid tested in this work ($128\times 128$),
the prototype already incurs higher estimated gate costs than the FT target,
though the difference remains \rev{below a factor of two}. 
Extrapolating to $256\times 256$ grids typical of high-resolution CTEM
imaging, the prototype's $O(N^2)$ scaling would require approximately
\rev{$3.3\times10^6$} T-gates compared to $\sim 6.1\times10^5$ for FT arithmetic,
representing a \rev{factor-of-five} disadvantage that would only worsen for
larger systems.

\subsection{Implications for \rev{fault-tolerant implementation} and future work}

The distinction between prototype and FT implementations clarifies the
development path toward \rev{fault-tolerant} CTEM simulation.
The prototype implementation successfully validates the quantum CTEM
algorithm against classical multislice simulations with exact numerical
agreement ($\rho = 1.000000$, Appendix~\ref{app:validation}) and enables
proof-of-concept hardware demonstrations on NISQ devices with limited
qubit counts (Appendix~\ref{app:ibm_hardware}). This validation establishes
that the physics-level quantum circuit-encompassing state preparation,
QFT operations, diagonal unitaries, and measurement---correctly implements
the POA propagation model independent of the specific gate synthesis
strategy employed.

\rev{Polylogarithmic per-run circuit scaling requires} transitioning to FT arithmetic circuits\rev{; as discussed in Sec.~\ref{sec:advantage}, this per-run scaling is necessary but not sufficient for an end-to-end advantage over classical FFT-based methods.} This transition will become practical as
error-corrected quantum processors with approximately 100 logical qubits
and fault-tolerant T-gate factories become available.\cite{fowler2012surface,litinski2019magic}
The resource estimates in Table~\ref{tab:resources} (Sec.~\ref{sec:resources})
reflect this FT target and indicate that total logical qubit requirements
remain modest, with approximately 58 qubits needed for $256\times 256$
grids and 70 for $1024\times 1024$ grids, even accounting for the ancilla
overhead required for arithmetic operations.

Intermediate strategies may emerge that combine classical preprocessing
for slowly varying components, such as precomputed atomic Gaussian
coefficients, with quantum arithmetic for grid-dependent evaluation,
such as coherent computation of squared distances $r^2$ and exponential
functions. Such hybrid schemes could reduce ancilla requirements while
maintaining subquadratic scaling for specific CTEM parameter regimes,
potentially offering practical advantages in the transition era between
NISQ and fully fault-tolerant quantum computing.

Both the prototype and FT implementations extend naturally to full
multislice propagation through alternating application of specimen and
propagation operators across multiple material slices. The prototype's
$O(N^2)$ cost per slice remains manageable for thin specimens requiring
approximately ten slices but becomes prohibitive for thick samples
demanding hundreds of propagation steps. The FT approach, with its
$O(\text{poly}(\log N))$ depth per slice, scales more favorably \rev{per run} to
arbitrary specimen thickness\rev{; whether this yields an end-to-end advantage for thick specimens depends on the output required and the number of repetitions (Sec.~\ref{sec:advantage}).}

In summary, the prototype implementation serves as a validated reference
platform for near-term quantum CTEM experiments, while the FT arithmetic
framework defines the asymptotic \rev{per-run} scaling target for fault-tolerant \rev{implementations}. The successful validation of the prototype against classical
simulations provides strong confidence that the FT implementation, when
realized on future error-corrected hardware, will inherit the same
numerical correctness demonstrated here\rev{, with the per-run gate-count scaling estimated above}. The physics-level agreement
is implementation-agnostic, depending only on the correct realization
of the underlying quantum operations (QFT, diagonal phases, measurement)
rather than the specific circuit synthesis strategy used to construct
those operations.

\end{document}